\documentclass[10pt]{iopart}
\usepackage{iopams}  
\expandafter\let\csname equation*\endcsname\relax
\expandafter\let\csname endequation*\endcsname\relax
\usepackage{amsmath, amsthm, amssymb}
\usepackage{graphicx} 
\usepackage[leftcaption]{./sidecap}
\usepackage{epsfig}
\usepackage{dcolumn}    % Align table columns on decimal point
\usepackage{xspace} % Sensible space treatment at end of simple macros

\newcommand{\vek}[1]{\boldsymbol{#1}}
\begin{document}
%%%%%%%%%

\title[Evolution of resonant binaries]{ Probing evolution of binaries influenced by the 
spin-orbit resonances
}

\author{A Gupta$^{1}$ \& A Gopakumar$^{1}$}
\address{$^1$ Department of Astronomy and Astrophysics, Tata Institute of Fundamental Research, Mumbai 400005,
India}
\ead{arg@tifr.res.in, gopu@tifr.res.in}

\begin{abstract}

 We evolve isolated comparable mass spinning compact binaries experiencing 
Schnittman's post-Newtonian spin-orbit resonances 
in an inertial frame associated with $\vek j_0$, the initial direction of the total
angular momentum.
We argue that accurate gravitational wave (GW)
measurements  of the initial orientations of the two
spins and orbital angular momentum from $\vek j_0$  should allow us to
distinguish between the two possible families of spin-orbit resonances.
Therefore, these measurements have the potential to provide
 direct observational evidence of possible binary formation scenarios.
The above statements should also apply for binaries that do not remain in a resonant plane 
when they become detectable by GW interferometers.
The resonant plane, characterized by the vanishing scalar triple product involving the two 
 spins and the orbital angular momentum, naturally appears in 
the one parameter family of equilibrium solutions, discovered by Schnittman.
%It turns out that inspiral templates for binaries residing in such resonant planes
%should be able to capture binaries under the influence of the spin-orbit resonances.
We develop a prescription to compute the time-domain inspiral templates
for binaries residing in these resonant configurations 
and explore their preliminary data analysis consequences.
% to support
% our deductions.

\end{abstract}

% 04.30.-w      Gravitational waves
% 04.25.Nx      Post-Newtonian approximation; perturbation theory; related approximations
% 97.60.Lf      Black holes
% 98.35.Jk      Galactic center, bar, circumnuclear matter, and bulge (including black hole and distance measurements)
%\pacs{04.30.--w, 04.25.Nx, 97.60.Lf, 98.35.Jk}

\submitto{\CQG}

% \maketitle

\section{Introduction}
Coalescing comparable mass compact binaries containing spinning stellar mass  
black holes (BHs) are among the expected gravitational wave (GW) sources for the ground based 
interferometric GW detectors like advanced LIGO (aLIGO), advanced Virgo and KAGRA \cite{Rana13}. 
In comparison, comparable mass spinning supermassive BH binaries will be required 
to realize GW astronomy in the milli-Hertz and nano-Hertz regimes in the coming decades \cite{KJLee11, PMS12}. 
The optimal data analysis method of {\it matched filtering} is being invoked
to extract the weak GW signals that are deeply buried in the noisy data sets.
In this method one cross-correlates the relevant data with 
several template banks that contain accurately modeled GW signals from a number of 
expected compact binary sources.
The construction of these GW search templates requires one to model 
GW polarization states, $h_{\times}(t)$ and
$h_{+}(t)$, associated with the inspiral phase of coalescing compact binaries
in an accurate and efficient manner.
 Fortunately, GWs emitted during the inspiral phase can be accurately modeled
by invoking the post-Newtonian (PN) approximation to general relativity.
In the case of non-spinning compact binaries, inspiraling along quasi-circular orbits,
theoretical inputs required to compute highly accurate inspiral phase templates
are available to the 3.5PN order.
These  include inputs to compute the fully 3.5PN accurate orbital phase 
evolution and 3PN accurate expressions for $h_{\times}(t)$ and
$h_{+}(t)$  ~\cite{BFIJ, BDFI, BF3PN}.
Recall that $n$PN corrections  provide contributions that are accurate to the relative  $(v/c)^{2n}$ order
beyond the ‘Newtonian’ estimate, where $v$ and $c$
are the orbital and light speeds, respectively.
Interestingly,  
there are on-going efforts to describe the compact binary dynamics at the conservative 4PN order 
~\cite{PS12}.
In the case of binaries containing 
Kerr BHs, the spin effects  should be incorporated 
while constructing appropriate search templates.
This is the main motivation for the ongoing efforts to compute the higher PN order corrections to the dominant
order spin-orbit and spin-spin contributions to the dynamics of spinning compact binaries, 
computed some four decades ago by  
Barker and O'Connell \cite{BO_75}.
We observe that for compact
binaries, containing maximally spinning BHs, the leading order spin-orbit and spin-spin 
interactions enter the orbital dynamics at 1.5PN and 2PN order, respectively \cite{LK_95}.
At present, the next-to-next-to-leading order spin-orbit contributions that appear at 3.5PN order are available in \cite{Marsat2013} while the 
next-to-next-to-leading order spin-orbit and spin(1)-spin(2) Hamiltonians are available in \cite{HSH2013}.
Very recently, 
the next-to-next-to-leading order spin-orbit contributions to the gravitational wave flux 
and associated orbital phase evolutions were obtained 
for binaries in quasi-circular orbits \cite{Bohe2013}.
%http://arxiv.org/abs/1303.7412
Strictly speaking, we only have all the relevant inputs to perform GW phasing to 2.5PN order 
while incorporating all the spin effects associated with the two maximally spinning 
Kerr BHs. In comparison, the ready-to-use amplitude corrected GW polarization states are only available to
2PN order for binaries in quasi-circular orbits \cite{ABFO,BFH,GG1} and to 1PN order for eccentric binaries ~\cite{GS11}.
In what follows, we explore the evolution of
certain binary configurations that contain 
two spinning compact objects of comparable  masses $m_1$ and $m_2$ ($m_1\gtrsim m_2$) 
and having Kerr parameters $\chi_1$ and $\chi_2$ such that their spin angular momenta
are given by $\vek S_1 = G\, m_1^2 \, \chi_1\, \vek s_1/c$ and  $\vek S_2 = G\, m_2^2 \, \chi_2\,\vek s_2/c$.

   Roughly a decade ago, Schnittman discovered certain equilibrium configurations for 
spinning and precessing compact binaries \cite{JS}.
The binaries in such configurations  have their two spins and the orbital
angular momentum $\vek L$ lying in the same plane and 
the definition of the total angular momentum,
namely $\vek J = \vek L + \vek S_1 + \vek S_2$, implies that 
$\vek J$ will also lie in this plane. 
 Schnittman observed that spinning compact binaries in these equilibrium configurations 
are characterized by constant values of $\vek s_1 \cdot \vek s_2$, $\vek k \cdot \vek s_1$ 
and $\vek k \cdot \vek s_2$ during the precessional time scale,
where $\vek s_1$, $\vek s_2$ and $\vek k$ are unit vectors along the two spins 
and orbital angular momenta, respectively.
Schnittman termed these equilibrium solutions as the ``spin-orbit resonant configurations"
as $\vek s_1$, $\vek s_2$ and $\vek L$ precess around $\vek J$ with a constant frequency
for such binaries in the absence of gravitational radiation reaction effects 
($\vek J$  is a conserved quantity both in magnitude and direction in the absence of 
GW damping).
With the inclusion of the reactive contributions in PN-accurate orbital dynamics, Schnittman observed that 
$\vek s_1 \cdot \vek s_2$ approaches unity on the inspiral timescale for 
certain type of equilibrium configurations.
Therefore,  the equilibrium configurations experience 
the spin alignment towards the end of their inspiral.
Binaries, not initially in the neighborhood of these equilibrium configurations,
can eventually get locked and librate around them during their inspiral.
The spin-orbit resonances may have important astrophysical implications as noted in 
~\cite{JS,KSB09,Gerosa13,BKS12}.
This is because of the ability of these resonances to align the spins 
of comparable mass supermassive BH binaries prior to their mergers \cite{KSB09,BKS12}.
This will ensure that the massive BHs formed via BH coalescences will not experience large recoil 
velocities. Therefore, the merger remnants may be retained in their host galaxies that are hierarchically
formed from the merger of smaller galaxies.
Very recently, it was argued that the BH spins in comparable mass stellar mass BH binaries
would lie preferentially in a resonant plane due
the spin-orbit resonances when their GWs enter the aLIGO frequency window \cite{Gerosa13}.
They demonstrated that PN evolution forces the two spins to lie
in certain resonant planes, characterized by either
$\Delta \tilde \phi = \pm 180^{\circ}$ or
$\Delta \tilde\phi = 0^{\circ}$ 
 where $\Delta \tilde \phi$ is the relative angular separation between
the two spins in an associated orbital plane
(this requires that the tides are efficient during the formation of these binaries).
The binaries belonging to these two
$\Delta \tilde \phi$ families are forced to belong to  what they termed as the standard mass ratio (SMR)
and the reverse mass ratio (RMR) binary formation scenarios.
In the SMR binary formation channel,
the more massive
star will evolve to form the more massive component of the BH binary and the compact binary during its inspiral
will be influenced by the $\Delta \tilde \phi = \pm 180^{\circ}$ spin-orbit resonances.
In contrast,  the heavier BH forms during  the second
supernova explosion in the RMR binary formation scenario and this is
essentially due to the substantial mass transfer during the Roche lobe overflow of
the progenitor.
Such BH binaries are expected to get influenced by the $\Delta \tilde \phi = 0^{\circ}$  resonances.
Gerosa {\it et. al} studied in detail the combined effects of
efficient tides and supernova kicks (both isotopic and polar)
on the above two binary formation channels. Their detailed explorations allowed them
to provide several distributions for the misalignments between $\vek k$ and the two spins
at orbital separations of roughly 500 Schwarzschild radius
(see the  $a=1000\,M$ scatter plots in figures~5 and 6 in \cite{Gerosa13}).
These scatter plots turned out to be very helpful in providing the initial conditions for our numerical
investigations.
 Gerosa {\it et. al} also explored the binary formation scenarios involving inefficient tides
and termed the resulting binaries as the freely precessing ones.
It was pointed out that the accurate matched filtering measurements of $\Delta \tilde \phi $
and $\theta_{12}=\cos^{-1}(\vek s_1 \cdot \vek s_2)$ from a large sample of
GW observations will constrain the various possible models of binary formation \cite{Gerosa13}.
We note that it is customary to probe the dynamics of these inspiraling and precessing binaries
in an orbital triad associated with
the Newtonian orbital angular momentum $\vek L_{\rm N} = \mu\, \vek r \times \vek v $,
where $\mu, \vek r$ and $\vek v$ are the reduced mass, orbital separation and 
velocity, respectively.

 In this paper, we evolve comparable mass spinning compact binaries
from an initial epoch characterized by $x= 10^{-3}$ (we usually denote this specific $x$ value as $x_{\rm i}$).
The 
dimensionless PN expansion parameter $x$ is defined 
in terms of an orbital-like frequency $\omega$ and the total mass $m$:
$x \equiv ( G\, m\, \omega /c^3)^{2/3}$.
We use the 
orbital angular momentum $\vek L$ rather than its Newtonian counterpart $\vek L_{\rm N}$ to describe the 
binary orbits.
%{\it This is influenced by the observation that 
%precessional equation, appropriate for $\vek L$, 
%is usually used to evolve $\vek L_{\rm N}$ while incorporating the effects of 
%the dominant order spin-orbit coupling in the literature \cite{GG1}.}
Additionally, we invoke an inertial frame associated with $\vek j_0$, the unit vector along the initial direction of 
the total angular momentum of the binary, 
to specify both the orbital and spin angular momenta.
In contrast, it is common to invoke an $\vek L_{\rm N}$-based non-inertial 
triad to specify the two spins at the initial epoch.
We observe that is customary to numerically evolve $\vek L_{\rm N}$ by invoking the
precessional equation appropriate for $\vek L$ while incorporating the effects of dominant 
order spin-orbit coupling \cite{GG1}.
The initial $x$ value
makes sure that these compact binaries inspiral essentially due to the 
 emission of GWs from orbital separations 
$\sim 500\,R_s$ where 
$R_s \equiv 2\, G\, m/c^2$ being the Schwarzschild radius. 
We terminate these numerical integrations when 
$x$ reaches either of the following two fiducial values: $x= x_0$ 
or $x = x_{\rm f} \equiv 0.1$.
In our numerical integrations,
$x_0 \equiv (G\, m\, \omega_0 /c^3)^{2/3} $ 
where 
$\omega_0 \sim 10\,\pi$ Hz for ground-based interferometers like aLIGO
while $ \omega_0 \sim 10^{-4}\, \pi$ Hz for eLISA.
 The $x_{\rm f}$ value is essentially influenced by the earlier 
investigations~\cite{JS,KSB09}.
We evolve comparable mass spinning compact binaries that 
 satisfy the set of equilibrium spin configurations at $x_{\rm i}$, discovered by Schnittman.
These  one parameter family of equilibrium  configurations,
characterized either by $\Delta \phi=0^{\circ}$ or $\pm 180^{\circ}$,
 can be obtained 
by demanding that both $\gamma\equiv [\vek k, \vek s_1, \vek s_2]=\vek k \cdot (\vek s_1 \times \vek s_2)$ and its time derivative 
should vanish at the initial epoch \cite{JS}. In our approach, $\Delta \phi$ provides the relative 
angular separation of the two spins in a plane perpendicular to $\vek j_0$.
Recall that these equilibrium configurations may be viewed as spin-orbit resonances
as the precession frequencies of $\vek L, \vek s_1$ and $\vek s_2$ around $\vek j_0$ are rather identical.
We argue that accurate matched filtering measurements of the orientations 
of $\vek s_1, \vek s_2$ and $\vek k$ from $\vek j_0$ at $x_0$
 should allow us to distinguish between binaries under the influence of either 
$\Delta \phi = 0^{\circ}$ or 
$\Delta \phi = \pm 180^{\circ}$ spin-orbit resonances.
Therefore, these accurate GW measurements from an inspiraling comparable mass spinning 
binary should provide, in principle, 
the direct observational evidence of 
 binary formation channels 
involving  the SMR or RMR scenarios that also involve efficient tides, as detailed in \cite{Gerosa13}.
To illustrate the above statement, let
$\theta_1', \theta_2'$ and $\iota'$ stand for 
the orientations of $\vek s_1, \vek s_2$ and $\vek k$ from $\vek j_0$ at $x_0$.
We show that the binaries that are influenced by $\Delta \phi = \pm 180^{\circ}$ resonances
tend to have $ \theta_1' > \theta_2' > \iota' $.
The typical  $\iota'$ values are usually lie below
$10^{\circ}$. However, negligible $\iota'$ values  
suggest very efficient tides during the binary formation.
In contrast, $\Delta \phi = 0^{\circ} $ resonant binaries tend to have 
$ \theta_2' > \theta_1' >  \iota' $ and
typical $\iota'$ values are $> 10^{\circ}$. 
Non-negligible $\iota'$ values in the range of few degrees indicate 
efficient tides during the binary formation.
We show that the above inferences 
also apply for binaries that do not remain in a resonant plane 
when they become detectable by GW interferometers.
The resonant plane, characterized by either $\Delta \phi = 0^{\circ}$ or
$\Delta \phi = \pm 180^{\circ}$ restrictions, 
naturally appears in the above mentioned  
one parameter family of equilibrium solutions.

 It turns out that the two black hole spins and the orbital angular momentum do not remain in a plane
during the late stages of inspiral 
for binaries that were not in Schnittman's equilibrium configurations at $x_{\rm i}$.
Indeed, these binaries are  influenced by the spin-orbit resonances
and get locked into a nearby resonant plane during their inspiral.
However, this may not be sufficient to force the above three vectors to share a common plane
when these binaries inspiral to $x_0$.
For such binaries,
the above listed angular variables librate around their resonant values
and the plots for $\Delta \phi (x)$ and $\gamma(x)$ can have non-negligible amplitudes 
during the late stages of inspiral.
We also emphasize the importance of measuring accurately
the values of $\theta_{12}$  at $x_0$.
The accurate $\theta_{12} (x_0)$ measurements turned out to be crucial to distinguish
the freely precessing binaries from those under the influence of the 
spin-orbit resonances.
Following Gerosa {\it et. al},
the freely precessing binaries are expected to have $\theta_1(x_{\rm i}) \sim \theta_2(x_{\rm i}) $
as the tidal interactions play no significant role during their formation.
Additionally, such binaries are not affected by the spin-orbit resonances during their 
inspiral from $x_{\rm i}$ to $x_0$ \cite{Gerosa13}.
% binaries inspiral from $x_{\rm i}$ to $x_0$ \cite{Gerosa13}.
% Additionally, the spin-orbit resonances essentially play no role as these
% binaries inspiral from $x_{\rm i}$ to $x_0$ \cite{Gerosa13}.
Our numerical integrations show that these binaries can mimic
the constraints on the $\theta_1', \theta_2'$ and $\iota'$ values
that are satisfied by the two resonant families.
However, the $\theta_{12} (x_0)$ values of freely precessing binaries
 will not obey two
specific relations, involving $\theta_1'$ and $\theta_2'$ values,
that are fulfilled by binaries affected by the spin-orbit resonances.
This is relevant as binaries under the influence of 
$\Delta \phi=0^{\circ}$ ($\Delta \phi=\pm 180^{\circ}$) spin-orbit resonances
are expected to have $\theta_{12} (x_0) \sim \theta_2' - \theta_1' $
($ \theta_{12} (x_0) \sim \theta_1' + \theta_2'$).
Therefore, the accurate measurements of $ \theta_1', \theta_2', \iota'$ and
$\theta_{12} (x_0)$ values are crucial to distinguish the three possible types
of inspiraling comparable mass spinning binaries.
These three possible types, as expected, include binaries that are either freely
precessing or influenced  by one of the two spin-orbit resonances (the $\Delta \phi = 0 $
or  $\Delta \phi = \pm 180^{\circ} $ resonances).

  We also develop  
a prescription to compute 
the time domain GW polarization states for comparable mass spinning 
compact binaries experiencing spin-orbit resonances in the aLIGO/eLISA frequency windows.
Our approach
invokes $\vek k$ to describe the binary orbits and
the $\vek j_0$-based inertial frame to specify the two spins and is based on \cite{GG1}.
Therefore, our approach can easily incorporate various expressions
that are required to analyze the spin-orbit resonances in a $\vek j_0$-based inertial frame.
We show that 
the temporally evolving $h_{\times, +}(t)$ are uniquely characterized by 
only six parameters at the fiducial 
 $x_0$ values for binaries that reside in the resonant planes.
These six parameters include the four  basic ones, namely
$m$, $\eta (=m_1\, m_2/m^2)$, $\chi_1$, $\chi_2$ and the two angular
parameters, $\theta_1 (x_0)$ and $\phi_1 (x_0)$, 
that specify the  orientation of more massive spin at $x_0$.
The requirement that $ \gamma $ and its time derivative should be zero 
at $x_0$ forces the orientation of $\vek s_2$ to become dependent parameters for such
binaries.
In comparison, one requires to specify eight parameters to obtain $h_{\times, +}(t)$ for binaries
not residing in the resonant plane.
This is essentially due to the non-vanishing  $ \gamma $ and $\dot \gamma$ values at $x_0$ for 
such binaries.
Invoking the match $\cal{M}$ computations, detailed in \cite{DIS98}, we compare inspiral templates for binaries residing 
in and librating around the resonant configurations in aLIGO frequency window. Binaries in 
`near-resonance' configurations tend to have $\cal{M}$ estimates $>0.9$ while the $\cal{M}$ 
estimates are $<0.9$ for binaries  in `far-resonance' configurations. The rather high 
$\cal{M}$ estimates point to the possibility that a computationally cheaper resonant inspiral template
bank may provide the desirable fitting factor $(FF)\gtrsim0.97$ for binaries influenced by spin-orbit resonances.
This is because $FFs$ are obtained by maximizing the $\cal {M}$ over all the templates present in a certain 
bank of inspiral waveforms.
% and these waveforms are expected to cover 
%It should be noted that the FF is obtained by maximizing the $\cal {M}$.... }

 The paper is organized in the following way.
In the next section, we briefly describe the spin-orbit resonances, detailed in ~\cite{JS}
and the way to analyze the spin-orbit resonances in the inertial frame associated with $\vek j_0$.
Various implications of our approach 
are probed in section~\ref{Sec_IIb}.
Our prescription to compute time-domain GW polarization states for inspiraling 
binaries experiencing the spin-orbit resonances is 
presented in section~\ref{Sec_III} along with certain preliminary data analysis implications. 
Conclusions are presented in section~\ref{Sec_dis_con}.

\section{ PN-accurate Equilibrium Configurations and their GW emission induced evolution}
\label{SecII}

We first summarize \cite{JS} that probed the evolution of 
 comparable mass precessing compact binaries initially residing in and around
 certain equilibrium spin configurations while invoking an orbital triad 
associated with $\vek L_{\rm N}$.
Section~\ref{Sec_IIb} contains our approach to describe the evolution of such binaries in 
 an inertial frame based on $\vek j_0$ along with various inferences.

\subsection{Spin-orbit resonances in an orbital triad}
\label{Sec_IIa}

  Schnittman invoked an orbital triad based on $\vek L_{\rm N}$ to describe 
the dynamics of comparable mass spinning compact
binary configurations as evident from figure~1 in \cite{JS}.
In what follows, we use an orbital triad based on  $\vek L$ rather than $\vek L_{\rm N}$
to describe these binaries.
For generic spinning compact binaries, the two spins are freely 
specified at the initial epoch by 
 four angles, namely
%freely choosing four angles
 $( \tilde \theta_{1}, \tilde \phi_{1})$ and
 $( \tilde \theta_{2}, \tilde \phi_{2})$.
Therefore, the unit vectors along the two spins read 
\begin{subequations} 
\label{Spin_s1_s2}
\begin{align}
\vek s_1(x_{\rm i}) &=  \sin \tilde \theta_{1}\,\cos \tilde \phi_{1} \, \vek a + \sin \tilde \theta_{1}\,\sin \tilde \phi_{1} \, \vek b 
+ \cos \tilde \theta_{1}\, \vek k \,,\\
\vek s_2(x_{\rm i}) &=  \sin \tilde \theta_{2}\,\cos \tilde \phi_{2} \, \vek a + \sin \tilde \theta_{2}\,\sin \tilde \phi_{2} \, \vek b 
+  \cos \tilde \theta_{2} \, \vek k \,,
\end{align}
\end{subequations}
where $\vek a$ and $\vek b $ may be identified with unit vectors $\vek e_{\rm x}$ and $\vek e_{\rm y}$ of \cite{JS}.
Additionally, Schnittman  equated  $\tilde \phi_1 $ 
at the initial epoch to zero by noting that
the orbital dynamics should be preserved under a rotation around $\vek k$.
This implies that the orientations of these binaries,
characterized by certain $m, \eta, \chi_1, \chi_2$ and $x$ (or $\omega$) values,
are specified by just {\it three} angular variables.
These variables are $(\tilde \theta_1, \tilde \theta_2, \Delta \tilde \phi = \tilde \phi_2 - \tilde \phi_1)$
where $\Delta \tilde \phi$ specifies the relative angular separation of the two spins in the orbital plane
while $\tilde \theta_1$ and $ \tilde \theta_2$ define the orientations of $\vek s_1$ and $\vek s_2$ from 
$\vek k$, respectively.
% as pointed out in ~\cite{KSB09}.
It is important to note that these angular variables vary over precessional and reactive time scales.
Further, it may be recalled that the dynamical evolutions of such binaries involve three time-scales associated 
with the orbital, precessional and inspiral aspects of their dynamics and we denote 
these timescales by $\tau_{\rm orb}, \tau_{\rm pre} $ and $\tau_{\rm rr} $, respectively.
It is not very difficult to infer that  $\tau_{\rm orb} \ll \tau_{\rm pre} \ll \tau_{\rm rr} $
as they are associated with the Newtonian, 1PN and 2.5PN order terms in the PN-accurate orbital dynamics.

  The equilibrium spin configurations, detailed in ~\cite{JS}, are obtained by demanding that the first 
and second derivatives of $\vek s_1 \cdot \vek s_2$ should be zero. Invoking the precessional equations
for $\vek s_1 $ and $ \vek s_2$, given by equations~(\ref{eq:s1_s2_dot}) below, it is easy to show that the above requirements are 
identical to equating $\gamma $ and its time derivative to zero \cite{JS}.
The expression for $\gamma$ in the orbital triad reads
\begin{align}
\label{eq:gamma_abk}
\gamma = \sin \tilde \theta_1 \,\sin \tilde \theta_2 \, \sin \Delta \tilde \phi \,, 
\end{align}
and the requirement that $\gamma =0$ implies that $\Delta \tilde \phi$ can take only one of 
the following two values: $0^{\circ}$ or $\pm 180^{\circ}$.
The constraints, namely $\gamma = \dot \gamma =0$, allow us to numerically obtain $\tilde \theta_2$ in terms 
of $\tilde \theta_1$ for a binary characterized by specific values of 
$ \Delta \tilde \phi, m$, $\eta, \chi_1$ and $\chi_2$. In other words, the  
solutions to the above two constraints trace out one-dimensional curves in ($\tilde \theta_1,\tilde \theta_2$) space \cite{JS}.
These solutions, having $\Delta \tilde \phi = 0^{\circ}$ or $\pm 180^{\circ}$, 
stand for specific configurations where $\vek k, \vek s_1$ and $ \vek s_2$ lie in a plane
such that $\vek L,  \vek s_1$ and $ \vek s_2$ precess around $\vek J$ with a roughly constant angular frequency
on a precessional time scale.
This prompted, as noted earlier,
 Schnittman to term these equilibrium configurations as certain spin-orbit resonant configurations.
The requirement that $\dot \gamma =0$ may be written as an algebraic constraint
invoking the precessional equations for $\vek k, \vek s_1$ and $\vek s_2$.
The constraint reads  
\begin{align}
\label{Eq_Eqm_equ}
(\vek \Omega_1 \times \vek S_1) \cdot [\vek S_2 \times (\vek L + \vek S_1)]=  
(\vek \Omega_2 \times \vek S_2) \cdot [\vek S_1 \times (\vek L + \vek S_2)] \,,
%\gamma = \sin \tilde \theta_1 \,\sin \tilde \theta_2 \, \sin \Delta \tilde \phi \,, 
\end{align}
where $\vek \Omega_1$ and $\vek \Omega_2$ provide precessional frequencies of $\vek s_1$ and $\vek s_2$.
 Schnittman incorporated the contributions arising from the leading order spin-orbit 
and spin-spin interactions in to the above precessional frequencies and these contributions may be extracted 
from \cite{BO_75,LK_95}.

 To probe the effect of gravitational radiation reaction on these equilibrium spin 
configurations, Schnittman wrote down PN-accurate differential equations for the following 
four  variables: $z_1=\vek k \cdot \vek s_1$, $z_2=\vek k \cdot \vek s_2$, 
$\beta=\vek s_1 \cdot \vek s_2$ and $\gamma = \vek k \cdot (\vek s_1 \times \vek s_2)$.
%, namely $z_1$, $z_2$, $\beta$ and $\gamma$.
It is straightforward to figure out that these differential equations arise from the PN-accurate 
precessional equations for $\vek k, \vek s_1$ and $ \vek s_2$ and therefore contain $x$.
This implies that the differential equation for $x$ that allows orbital frequency to slowly vary over 
$\tau_{\rm rr}$ may be invoked to incorporate the effects of gravitational radiation reaction
on these  variables. Therefore,
the differential equations required to describe the dynamics of precessing compact binaries spiraling in from
$x=10^{-3}$ in an orbital triad read
\begin{subequations}
\label{eq:s1_s2_dot}
\begin{eqnarray}
\dot z_1  &=& \frac{c^3}{G\, m}\, x^3 \, \gamma \, \chi_2 \, \biggl \{\frac{\delta_2}{q}-\frac{X_2^2}{2}-\frac{3}{2}\, x^{1/2}\, \eta\, \chi_1\, z_1 \biggr \}         \,,\\
\dot z_2  &=& \frac{c^3}{G\, m}\, x^3 \, \gamma \, \chi_1 \, \biggl \{-\delta_2\,q+\frac{X_1^2}{2}+\frac{3}{2}\, x^{1/2}\, \eta\, \chi_2\, z_2 \biggr \}        \,,\\
\dot \beta &=& \frac{3}{2}\, \frac{c^3}{G\, m}\, x^{5/2}\, \gamma \, \biggl \{X_2-X_1+x^{1/2} \, (X_1^2 \, \chi_1\, z_1  
              -X_2^2 \, \chi_2\, z_2 ) \biggr \}       \,,\\
\label{eq:gammadot_abk}
 \dot{\gamma} &=& \frac{c^3}{G\, m}\, x^{5/2} \, \bigg \{ \frac{3}{2}\, \frac{\delta m}{m}\, (\beta -z_1\, z_2)   
 + x^{1/2}\,\biggl [\delta_1\, q\, \chi_1\, (z_2-\beta\, z_1)  \nonumber \\
 &&+ \frac{\delta_2}{q}\, \chi_2\, (\beta\, z_2 - z_1)  
 + \frac{1}{2}\, X_1^2\, \chi_1\, (-z_2 -2\, \beta\, z_1 + 3\, z_1^2\, z_2)  \nonumber \\
 &&+ \frac{1}{2}\, X_2^2\, \chi_2\, (z_1 +2\, \beta\, z_2 - 3\, z_1\, z_2^2) \biggr ] 
 + \frac{3}{2}\, x \, \eta \, \chi_1 \, \chi_2 (z_1^2-z_2^2) \bigg \}\,,   \\              
\dot x &=& \frac{64}{5}\,\frac{c^3}{G\, m}\, \eta \, x^5 \,, 
\end{eqnarray}
\end{subequations}
 where $z_1, z_2$ and $\beta$ 
are specified by the angles 
$\tilde \theta_1, \tilde \theta_2 $ and $\theta_{12}$, respectively and 
we have verified that these equations are comparable to equations~(A10) in \cite{JS}.
In the above equations, 
$q, \delta m, X_1$ and $X_2$ stand for 
$m_1/m_2, (m_1-m_2), m_1/m$ and $m_2/m$, respectively while 
$\delta_{1, 2}=\eta/2 + 3 \left(1 \mp \sqrt{1 - 4 \eta} \right)/4$.
The presence of $x$ in the above expression indicates that these spin-orbit resonances
can sweep through a substantial portion of the  $(\tilde \theta_1, \tilde \theta_2)$ space 
during  the GW emission induced inspiral.
The fact that $\Delta \tilde \phi$ of a generic spinning compact binary 
can vary over the precessional timescale $\tau_{\rm pre}$ implies that 
the binary may approach the resonant values, $\Delta \tilde \phi = 0^{\circ}$ or $\pm 180^{\circ}$,
at some point during its lengthy PN-accurate inspiral regime, characterized by $\tau_{\rm rr}$.
It was argued that 
the orbital evolution of such generic spinning compact binaries will be heavily influenced by these
spin-orbit resonances \cite{JS}.
In practice, it is convenient to  
 numerically solve the following {\it four} differential equations, namely 
$ d z_1/dx =  \dot z_1/\dot x,  d z_2/dx =  \dot z_2/\dot x, d \beta/dx =  \dot \beta /\dot x$ and
$ d \gamma/dx =  \dot \gamma/\dot x$ to probe how these binary configurations evolve
under the combined influences of precessional and reactive dynamics from $x=10^{-3}$ to 
the late stages of inspiral prior to their coalescence.

 Numerical integration of the above equations allowed ~\cite{JS} 
to infer that the initial equilibrium configurations, defined by $\gamma = \dot \gamma =0$ at $x= x_{\rm i} \equiv 10^{-3}$,
remain in their resonant plane during the inspiral regime. 
Moreover, gravitational radiation reaction forces binary spins,
 initially not in the resonant plane,
to get locked and then librate about the equilibrium configurations
during its inspiral 
from $x_{\rm i}$ to $x_{\rm f} $ $(x=0.1)$.
This was  demonstrated by
showing that the instantaneous phase difference $\Delta \tilde \phi$ oscillates around $0^{\circ}$ with steadily decreasing
amplitude as evident from figure~5 in ~\cite{JS}.
This spin alignment prompted Kesden {\em et al.}~\cite{KSB09} to re-analyze these spin-orbit resonances 
in great detail and explore its implications for merging spinning BH binaries.
Very recently, it was pointed out that the BH spins in comparable mass stellar mass BH binaries 
would preferentially lie in a resonant plane, characterized by 
$\Delta \tilde \phi = 0^{\circ}$ or 
$\Delta \tilde \phi = \pm 180^{\circ}$, when GWs from such binaries enter the aLIGO frequency window \cite{Gerosa13}.
The above conclusion requires an admissible assumption that the spins of the BH progenitors 
should be partially aligned with the orbital angular momentum due to efficient tidal interactions.
The ability of such binaries to stay essentially in a resonant plane is due to the combined effects of spin-orbit resonances
and GW emission induced inspiral \cite{Gerosa13}. 
The authors also stated that it will be desirable to construct templates for inspiraling binaries influenced
by the spin-orbit resonances.

   In the next subsection, we evolve comparable mass binaries having 
spin configurations at $x=10^{-3}$ that are influenced by Schnittman's one parameter family of equilibrium solutions.
We invoke an inertial frame associated with $\vek j_0$
to specify 
both the orbital and spin angular momenta of such binaries
and to describe their PN-accurate evolution.
This is partly influenced by the observation that temporally evolving $h_{\times,+}(t)$, associated with spinning
compact binaries, are usually computed in such an inertial frame.
Therefore,  the various inputs that are required to describe the  spin-orbit resonances in the $\vek j_0$  frame
may be invoked while constructing 
inspiral templates associated with such binaries.

\begin{figure}[!ht]
\begin{center}
\includegraphics[width=75mm,height=75mm]{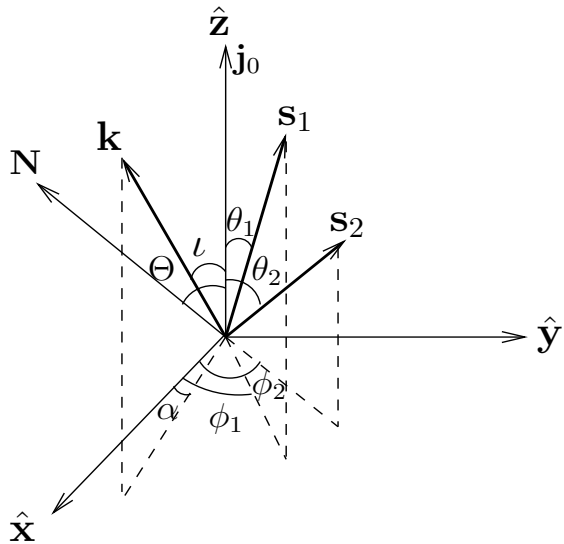}
\end{center}
\caption{The inertial frame based on $\vek j_0$, the unit vector along the total angular momentum at the initial epoch.
We let $\vek j_0$ to point along the $z-$axis while the line of sight vector $\vek N$ is defined by the constant
angle $\Theta$. The angles that specify the unit vectors along the orbital and spin angular momenta, denoted 
by $\vek k, \vek s_1$ and $\vek s_2$, are displayed. We depict the projections of these unit vectors onto the $x-y$ plane by dashed lines.
}
\label{figure:frame}
\end{figure}

\subsection{Spin-orbit resonances in an inertial frame defined by $\vek j_0$}
\label{Sec_IIb}

 We begin by describing how we specify the generic comparable mass spinning compact binaries,
characterized by specific values of $m, \eta, \chi_1 $ and $\chi_2$,
in an inertial  frame associated with $\vek j_0$ at large orbital 
 separations ($r \sim 500R_{\rm s} $ or $x=10^{-3}$).
%, in the invariant frame associated with $\vek j_0$ and $\vek N$.
We invoke the Cartesian coordinate system associated with this inertial frame such that the unit vectors 
along $\vek S_1, \vek S_2$ and $\vek L$ have the following $(\hat {\vek x},\hat {\vek y},\hat {\vek z})$ components:
\begin{subequations}
\begin{align}
% \vek k &= (\sin \iota\,\cos \alpha\, ,\,\sin \iota\,\sin
% \alpha\, ,\,\cos \iota)\,,\\
\label{eq:s1_s2_1}
\vek s_1 &= \left ( \sin \theta_1\,\cos \phi_1, \sin \theta_1\,\sin \phi_1, \cos \theta_1 \right )\,,\\  
\label{eq:s1_s2_2}
\vek s_2 &= \left ( \sin \theta_2\,\cos \phi_2, \sin \theta_2\,\sin \phi_2, \cos \theta_2 \right )\,,\\
\label{eq_k}
\vek k &= (\sin \iota\,\cos \alpha\, ,\,\sin \iota\,\sin
 \alpha\, ,\,\cos \iota)\,.
\end{align}
\end{subequations}
Therefore, it appears that we will require six angles, namely $\theta_1$, $\phi_1$, $\theta_2$, $\phi_2$, $\iota$ and 
$\alpha$, to specify the orientation of our binary 
in the invariant frame as displayed in figure~\ref{figure:frame}.
However, the fact that the invariant frame is defined such that the total angular momentum at the initial 
epoch points along the $z$-axis allows us to estimate the initial $x$ and $y$ components of $\vek k$
in terms of $m$, $\eta$, $\chi_1$, $\chi_2, x_{\rm i} $
and 
% $\theta'_1$, $\phi'_1$, $\theta'_2$, $\phi'_2$, 
the values of $\theta_1$, $\phi_1$, $\theta_2$ and $\phi_2$ at the initial epoch.
In other words, the initial $x$ and $y$ components of $\vek k$ become dependent variables 
as we equate the $x$ and $y$ components of $\vek j_0$ to zero. The resulting expression for the
initial  $k_x$ and $k_y$ read
\begin{subequations}
\label{Eq_ialpha_ini}
% \begin{eqnarray}
\begin{align} 
k_{\rm x}(x_{\rm i})=  \sin \iota\, \cos \alpha&=&-\frac{G\, m^2}{c\,L_{\rm i}}\,
\{X_1^2\, \chi_1\,\sin \theta_1\,\cos \phi_1  
+X_2^2\, \chi_2\,\sin \theta_2\,\cos \phi_2\} \,, \\
k_{\rm y}(x_{\rm i})= \sin \iota \sin \alpha&=&-\frac{G\, m^2}{c\,L_{\rm i}}\,
\{X_1^2\, \chi_1\,\sin \theta_1\,\sin \phi_1   
+X_2^2\, \chi_2\,\sin \theta_2\,\sin \phi_2\} \,, 
%k_{z,0}&=&+\sqrt{1-k_{x,0}^2-k_{y,0}^2} \,,
\end{align}
% \end{eqnarray}
\end{subequations}
where we employ 
the Newtonian accurate expression for 
 $|\vek L|$ at $x_{\rm i}$, given by
$ L_{\rm i}= G\, m^2\, \eta/(c\, \sqrt{x_{\rm i}})$. Additionally, the initial value of $k_z$ is uniquely
given by  $k_z=+\sqrt{1-k_x^2-k_y^2}$  as $\vek L$ almost point along $\vek J$ at $x_{\rm i}$. Therefore, 
the initial orientation of the orbital angular momentum
% , namely $\iota'$ and $\alpha'$, 
are uniquely 
given by equations~(\ref{Eq_ialpha_ini}). 
This implies that the dynamics of such binaries are uniquely obtained by freely 
specifying the initial values of the four angular variables that provide 
the orientations of the two spins in the invariant frame.
% namely $\theta'_1$, $\phi'_1$, $\theta'_2$ and $\phi'_2$.

   We invoke the following set of three vectorial and one scalar differential equations to 
describe the inspiral dynamics of precessing spinning compact binaries, extracted from
~\cite{BO_75,JS92,Racine_08}. We first list the precessional equations for $\vek s_1$ and $\vek s_2$,
given by
\begin{subequations}
\label{eq:s1_s2_dot}
\begin{eqnarray}
%\begin{align}
{\dot {\vek s}_{1}} &=&  \frac{c^3}{Gm}\, x^{5/2}\,
\Bigg\{\delta_1 \left(\vek k\times \vek s_1\right)  
+ \frac{1}{2}\,x^{1/2}\, \Big[ X_2^2\,\chi_2\,(\vek s_2 \times \vek s_1)    \nonumber \\  
&&-3\, X_2^2\,\chi_2\,(\vek k \cdot \vek s_2) \, (\vek k \times \vek s_1) -3\, \eta\, \chi_1\, (\vek k \cdot \vek s_1)\, (\vek k \times \vek s_1)\Big] \Bigg\} \,, \\
%%%%%%%%%%%%%%%%%%%%%%%%%%%%%%%%%%%%%%%%%%%%%%%%%%%%%%%%%%%%%%%%%%%%%%%%%%%%%%%%%%%%%%%%%%%%%%%%%
{\dot {\vek s}_{2}} &=&  \frac{c^3}{Gm}\,x^{5/2}\,
\Bigg\{\delta_2 \left(\vek k\times \vek s_2\right)  
+ \frac{1}{2}\,x^{1/2}\,\Big[  X_1^2\,\chi_1\,(\vek s_1 \times \vek s_2)   \nonumber \\
&&-3\,  X_1^2\,\chi_1\,(\vek k \cdot \vek s_1) \, (\vek k \times \vek s_2) -3\, \eta\, \chi_2\, (\vek k \cdot \vek s_2)\, (\vek k \times \vek s_2)\Big] \Bigg\} \,,
\end{eqnarray}
\end{subequations}
where the terms proportional to $x^{5/2}$ and $x^3$ incorporate the dominant order spin-orbit
and spin-spin interactions, respectively, for binaries moving in circular orbits.
The  $x^3$ terms that are proportional to $\eta$ are due to the quadrupole-monopole self interaction \cite{Racine_08}.
These terms were not included in the original analysis of Schnittman while they are present in 
the detailed analysis of 
\cite{KSB09}.
The precessional motion of $\vek L$ is described by
\begin{eqnarray}
\label{eq:kdot}
{\dot {\vek k}} &=&  \frac{c^3}{Gm}\, x^{3}\,\Bigg\{ \delta_1\,q\, \chi_1\, 
\left( \vek s_1\times\vek k\right)  
+\frac{\delta_2}{q}\, \chi_2\, \left( \vek s_2\times\vek k\right)  \nonumber \\
&&-\frac{3}{2}\,x^{1/2}\, \Bigl[ \eta\, \chi_1\, \chi_2\,(\vek k \cdot \vek s_1)\, (\vek s_2 \times \vek k) 
+  \eta\, \chi_1\, \chi_2\,(\vek k \cdot \vek s_2 )\, (\vek s_1 \times \vek k)       \nonumber \\
&&+ X_1^2\, \chi_1^2\,(\vek k \cdot \vek s_1)\, (\vek s_1 \times \vek k)
+  X_2^2\, \chi_2^2\,(\vek k \cdot \vek s_2)\, (\vek s_2 \times \vek k)\Bigr]\Bigg\} \,, 
\end{eqnarray}
and this  equation arises from the conservation of total angular momentum which leads to
 $\dot{\vek L}=- S_1 \, \dot{\vek s_1}-S_2\, \dot{\vek s_2}$.
%, where $\vek S_1=S_1\vek s_1$ and$\vek S_2=S_2 \vek s_2$.
 We incorporated the effects of gravitational radiation reaction
which causes the binaries to inspiral from $x_{\rm i}$ via
% \begin{subequations}
% \begin{eqnarray}
\begin{equation}
\label{Eq_xdot}
 \frac{ d x}{dt} = \frac{64}{5}\frac{c^3}{Gm}\eta\, {x}^5 \,.
 \end{equation}
% \end{eqnarray}
% \end{subequations}
This equation requires the quadrupolar order GW luminosity along with 
the energy balance argument \cite{JS92, PM}.
We have verified that our inferences about the values of various angular variables
at $x_0$ are rather insensitive to 
the inclusion of PN corrections to the above expression for
$\dot{x}$ .

To describe PN-accurate evolution of comparable mass spinning compact binaries, we
employ the Cartesian components of the precessional equations while 
numerically integrating equations~(\ref{eq:s1_s2_dot}), (\ref{eq:kdot}) and (\ref{Eq_xdot}).
In practice, we numerically solve the nine Cartesian components of the following three 
equations, namely  $d \vek k /dx, d \vek s_1 /dx $ and  $d \vek s_2 /dx$ 
where, for example, $ d \vek k /dx=(d \vek k /dt)/ \dot{x}$. 
This is how we follow the orientations of the two spins 
and the orbital angular momenta
 from $\vek j_0$  as these binaries inspiral from
$x_{\rm i}$ to $x_{\rm f}$. 
During our numerical integrations, the values of  
$\theta_1, \phi_1, \theta_2, \phi_2, \iota $
and $\alpha $ are extracted at the 
stipulated $x$ values from 
the three Cartesian components
of $\vek s_1, \vek s_2$ and $\vek k$
(we invoke equations~(\ref{Eq_ialpha_ini}) only once to estimate the initial Cartesian components of $\vek k$).
For example, 
the angular variables of the dominant spin, namely $ \theta_1$ and $\phi_1$,
are obtained via $\theta_1 = \cos^{-1} (s_{\rm 1z})$ and
$ \phi_1 = \tan^{-1}(s_{\rm 1y}/s_{\rm 1x}) $.
Similar 
 expressions are employed to obtain $ \theta_2$ and $\phi_2$ values while 
the orbital inclination $\iota$ and $\alpha$ values are also uniquely
extracted from the three Cartesian components of $\vek k$.
We are now in a position to explore PN evolution of binaries having
spin configurations, at $x=10^{-3}$, that are influenced by Schnittman's one parameter family of equilibrium solutions.

 We begin by listing 
the expression for  $\gamma$ in terms of angular variables defined in the inertial frame:
\begin{eqnarray}
 \label{eq:gamma}
 \gamma &=& \sin \theta_1 \, \cos \theta_2  \, \sin \iota \, \sin (\phi_1-\alpha)  
 - \sin \theta_2 \, \cos \theta_1 \, \sin \iota \, \sin (\phi_2 - \alpha) \nonumber \\
 &&- \sin \theta_1\, \sin \theta_2 \, \cos \iota \, \sin (\phi_1-\phi_2)\,.
 \end{eqnarray}
The fact that the equilibrium configurations are characterized by $\gamma=0$ implies 
%$\gamma $ should be zero for Schnittman equilibrium configurations demands that 
$\sin(\phi_1-\phi_2)=\sin(\phi_1-\alpha)=\sin(\phi_2-\alpha)=0$ for such configurations.
%as evident from the above equation.
Therefore, these special configurations
are also characterized by $\Delta \phi= \phi_2 - \phi_1 = 0^{\circ}$ or $ \pm 180^{\circ}$ 
in our inertial frame. 
%similar to Ref.~\cite{JS}.
Note that $\gamma =0$ forces certain restriction on the initial $\alpha$ value
and it turned out to be consistent with initial $\alpha$ value  
via equations~(\ref{Eq_ialpha_ini}).
For numerically obtaining Schnittman equilibrium configurations in the inertial frame, 
we employ the following expression for $\gamma$ involving the Cartesian components of $\vek k$
\begin{eqnarray}
 \label{eq:gamma_ifc}
 \gamma &=& k_{x} (\sin \theta_1 \sin \phi_1 \cos \theta_2-\sin \theta_2 \sin \phi_2 \cos \theta_1)  
     + k_y (\sin \theta_2 \cos \phi_2  \cos \theta_1  \nonumber \\
     &&- \sin \theta_1 \cos \phi_1  \cos \theta_2)    
     + k_z \sin \theta_1  \sin \theta_2  \sin (\phi_2-\phi_1)  \,,
 \end{eqnarray}
where we express $k_x$, $k_y$ and $k_z $ in terms of the Cartesian components 
of $\vek s_1$ and $\vek s_2$ invoking equations~(\ref{Eq_ialpha_ini}).
The associated expression for $\dot{\gamma}$
%, required to construct the equilibrium configurations in the invariant frame,
 is given by

\begin{eqnarray}
\label{eq:gammadot}
 \dot{\gamma} &=& \frac{c^3}{G\, m}\, x^{5/2} \, \bigg \{ \frac{3}{2}\, \frac{\delta m}{m}\, (\beta -z_1\, z_2) 
 + x^{1/2}\,\biggl [\delta_1\, q\, \chi_1\, (z_2-\beta\, z_1)   \nonumber \\
 && + \frac{\delta_2}{q}\, \chi_2\, (\beta\, z_2 - z_1)  
 + \frac{1}{2}\, X_1^2\, \chi_1\, (-z_2 -2\, \beta\, z_1 + 3\, z_1^2\, z_2)  \nonumber \\
 &&+ \frac{1}{2}\, X_2^2\, \chi_2\, (z_1 +2\, \beta\, z_2 - 3\, z_1\, z_2^2) + \frac{3}{2}\, \eta\, (\beta -z_1\, z_2)\, (\chi_1\, z_1 -\chi_2 \, z_2) \biggr ]   \nonumber \\
 &&+ \frac{3}{2}\, x \, \biggl[ \eta \, \chi_1 \, \chi_2 (z_1^2-z_2^2)   
 + X_1^2\, \chi_1^2\, z_1 \, (z_1\, \beta -z_2)      \nonumber \\
 &&+ X_2^2\, \chi_2^2\, z_2 \, (z_1 -\beta\, z_2) \biggr] \bigg \}\,,
\end{eqnarray}
where, $z_1$, $z_2$ and $\beta$ are functions of $\theta_1$, $\phi_1$, $\theta_2$, $\phi_2$. These 
three dot products in the invariant frame may be written as 
\begin{subequations}
\label{eq_z1_z2_b_xyz}
\begin{eqnarray}
 z_1 &=& \vek k \cdot \vek s_1 = k_x\, \sin \theta_1 \, \cos \phi_1 + k_y \, \sin \theta_1 \, \sin \phi_1 + k_z \, \cos \theta_1 \, \\
 z_2 &=& \vek k \cdot \vek s_2 = k_x\, \sin \theta_2 \, \cos \phi_2 + k_y \, \sin \theta_2 \, \sin \phi_2 + k_z \, \cos \theta_2 \, \\
 \label{Eq_beta}
 \beta &=& \vek s_1 \cdot \vek s_2=\sin \theta_1 \, \sin \theta_2\, \cos (\phi_1 -\phi_2) + \cos \theta_1 \, \cos \theta_2\,.
\end{eqnarray}
\end{subequations}
% \end{widetext}
%
It should be noted that the above expression for $\dot{\gamma}$ is not identical to equation ~(\ref{eq:gammadot_abk})
and this is due to the use of fully 2PN-accurate precessional equations that include 
the quadrupole-monopole self 
interaction terms in this subsection \cite{Racine_08} .
Let us note that we need to express 
$k_x$, $k_y$ and $k_z $, appearing in the above expressions for $z_1$ and $z_2$,
 in terms of the Cartesian components 
of $\vek s_1$ and $\vek s_2$ with the help of equations~(\ref{Eq_ialpha_ini})
while dealing with the above equation for $\dot \gamma$.
We obtain 
 Schnittman's equilibrium configurations  by simultaneously equating 
equations~(\ref{eq:gamma_ifc}) and (\ref{eq:gammadot}) for  $\gamma$ and $\dot \gamma $ to zero.
This allows us to numerically 
obtain $\theta_2$ values in terms of $(\theta_1, \phi_1)$ for two specific
values of $\Delta \phi$, namely $\Delta \phi=0^{\circ}$ or $\pm 180^{\circ}$
and define the equilibrium configurations in the invariant frame.
These configurations are such that initial $\alpha$ value, obtained via equations~(\ref{Eq_ialpha_ini}),
is consistent with the requirement arising from 
 equating $\gamma$, given by equation~(\ref{eq:gamma}), to zero.
Additionally, the extracted values of the angular variables are also consistent
with the requirement that the initial $x$ and $y$ components of $\vek J$ should be zero.
It should be noted that 
in our approach 
Schnittman's equilibrium configurations are specified by three angular variables, namely ($\theta_1$, $\phi_1$, $\Delta \phi$),
 compared to  two in ~\cite{JS}.
This is merely a consequence of invoking the Cartesian components to obtain these special configurations
and we have verified that our results do not depend on the initial value of $\phi_1$.
In what follows, we consider maximally spinning BH binaires with $m=20M_{\odot}$ and $q=11/9$
while choosing the initial $\phi_1$ to be $45^{\circ}$.
We are now in a position to explore and evolve binaries having 
spin configurations that are influenced by Schnittman's one parameter family of equilibrium solutions.

\begin{figure}[!ht]
\begin{center}
$\begin{array}{cc}
\includegraphics[width=65mm,height=60mm]{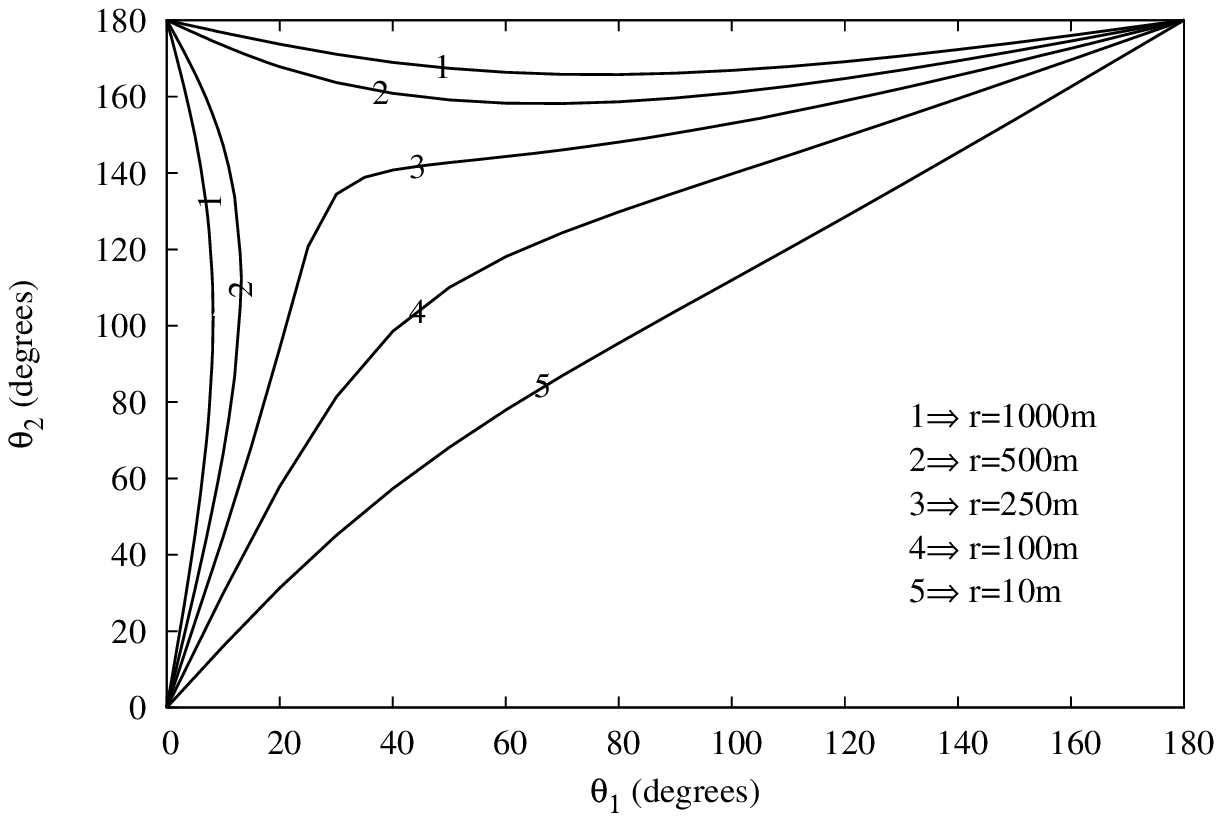}&
\includegraphics[width=65mm,height=60mm]{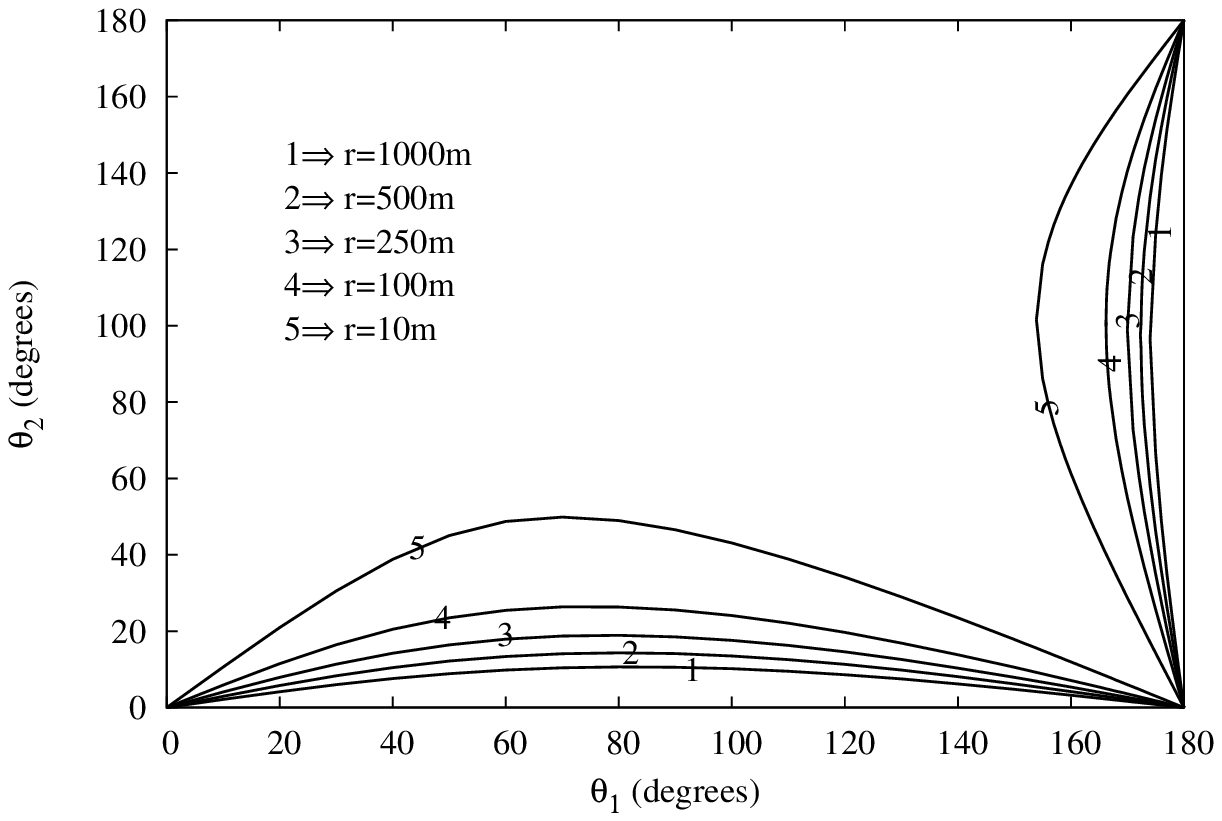}
\end{array}$
\end{center}
\caption{ 
Schnittman's equilibrium solutions in the $\vek j_0$-based inertial frame.
These curves are obtained by 
 equating the equations~(\ref{eq:gamma_ifc}) and ~(\ref{eq:gammadot}) to zero for maximally spinning
BH binaries with $q=11/9$.
The resonant configurations with $\Delta \phi=0^{\circ}$ are displayed on the left
panel while the right panel plots are for the $\Delta \phi=180^{\circ}$ equilibrium configurations.
}
\label{figure:th1_th2}
\end{figure}

 We begin by displaying Schnittman's equilibrium configurations in our inertial frame 
for maximally spinning BH binaries with  $q=11/9$ as 
one dimensional curves in the ($\theta_1$, $\theta_2$) plane (see figure~\ref{figure:th1_th2}).
These curves, influenced by figures~2 and 3 in ~\cite{JS}, 
are for binaries residing at various orbital separations starting from $r \sim 500\, R_s$.
The equilibrium solutions having $\Delta \phi =0^{\circ}$ are displayed in the left panel 
and the right panel plots are for $\Delta \phi =180^{\circ}$ equilibrium configurations.
These plots are fairly similar to the $\vek k \cdot \vek s_1$ - $\vek k \cdot \vek s_2$ plots in figure~1 of \cite{KSB09}
that depict Schnittman's equilibrium configurations in an orbital triad.
We observe that $\Delta \phi =0^{\circ}$ resonances have $\theta_2 > \theta_1$ and therefore the associated 
plots appear above the $\theta_2 = \theta_1$ diagonal while the reverse holds good for the 
$\Delta \phi =180^{\circ}$ equilibrium configurations.
For binaries having $\theta_1 < 10^{\circ}$ at large orbital separations like $r \sim 500\, R_s$,
the $\Delta \phi =0^{\circ}$
equilibrium solutions lie along two curves. This allows many more spin configurations to lie in the close
neighborhood of such resonant configurations.
Similar statement apply for binaries having $\theta_1(x_{\rm i}) > 160^{\circ}$ while dealing with 
$\Delta \phi =180^{\circ}$ equilibrium configurations.
We also observe that as these binaries inspiral from $x=10^{-3}$, the  $\Delta \phi =0^{\circ}$ equilibrium solutions 
sweep through a larger area of the ($\theta_1$, $\theta_2$) plane compared to the
$\Delta \phi =180^{\circ}$ configurations.
 All these conclusions are consistent with the studies that invoked $\vek L_{\rm N}$-based non-inertial orbital triad
to describe these  equilibrium solutions \cite{JS,KSB09}.

\begin{figure}[!ht]
\begin{center}
\includegraphics[width=115mm,height=40mm]{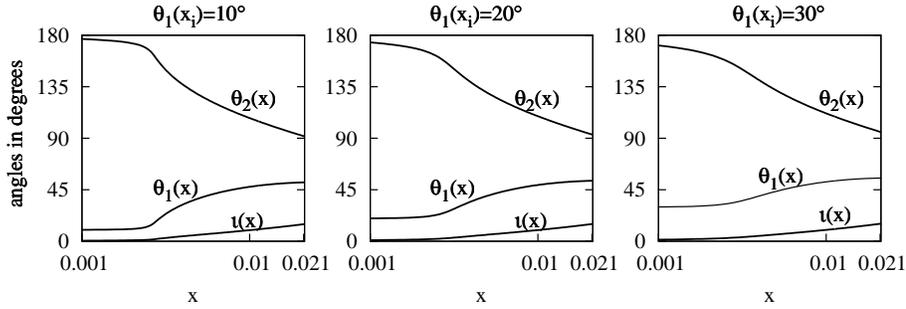}
\end{center}
\caption{ 
Semi-log plots that follow the GW induced evolution of the orientations of 
$\vek s_1, \vek s_2$ and $\vek k$ from $\vek j_0$ for our typical BH binaries.
We follow the evolution in the $[x_{\rm i}-x_0]$ interval for three spin configurations
that satisfy Schnittman's $\Delta \phi=0^{\circ}$ equilibrium solutions at $x_{\rm i}$.
It should be evident that $ \theta_2 > \theta_1 > \iota$ at $x_0$ and these angular
variables evolve secularly in the $x$ interval.
}
\label{figure:Res_th1_th2_i_th12}
\end{figure}

   We move on to probe the effect of GW induced damping on these equilibrium configurations.
The plots of figure~\ref{figure:Res_th1_th2_i_th12} probe the evolution of 
three  $\Delta \phi =0^{\circ}$ equilibrium solutions while terminating the inspiral at $x_0$.
These configurations are characterized by 
three values of $\theta_1 (x_{\rm i})$, namely  $10^{\circ}, 20^{\circ} $ and $30^{\circ} $.
The $\theta_1(x_{\rm i})$ values are influenced by the inference that the traditional formation scenarios 
for comparable mass BH binaries likely to result in spin-orbit misalignments $\leq 30^{\circ}$ \cite{Kalogera}.
%These values are also consistent with the green dots in the $a= 1000\, M$ scatter plot in figure 8 of \cite{Gerosa13}
%that detail binary formation models involving inefficient tides and polar supernovae kicks.
Note that this investigation actually provided estimates for  
 spin-orbit misalignments, namely $\tilde \theta_1$ values,
at large orbital separations.
The large orbital separations allowed us to 
let $\tilde \theta_1 (x_{\rm i}) \sim \theta_1 (x_{\rm i})$.
Secular increments/decrements in the $\theta_1(x)/\theta_2(x)$ values are clearly visible
while $\iota$ experiences secular increase during the inspiral.
It turned out that the secular evolution of various angular variables is a characteristic 
of binaries lying in the resonant planes, specified by either $\Delta \phi =0^{\circ}$
or $\Delta \phi = \pm 180^{\circ}$.
More importantly,
%The dashed vertical line in these plots allows us to read off the values of $\theta_1, \theta_2$ and 
%$\iota$ at $x_0$.
we find that $ \theta_2 > \theta_1 > \iota$ at the initial frequencies of GW detectors
and typical values of $\iota(x_0)$ are $ > 10^{\circ}$.

\begin{figure}[!ht]
\begin{center}
\includegraphics[width=115mm,height=40mm]{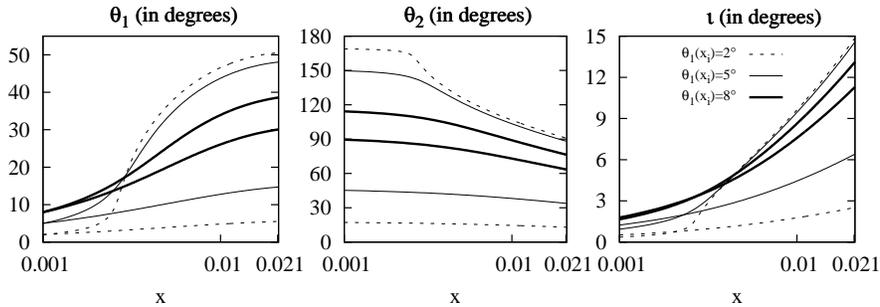}
\end{center}
\caption{ 
Plots for $\theta_1(x), \theta_2(x)$ and $\iota(x)$ associated with 
the $\Delta \phi=0^{\circ}$ equilibrium solutions for three 
$\vek s_1$ orientations that are $\leq 10^{\circ}$ at $x_{\rm i}$.
For a given $\theta_1(x_{\rm i})$ value, these spin configurations can have 
multiple  resonant $\theta_2(x_{\rm i})$ values.
This may be attributed  
to the presence of two equilibrium curves
at $r=1000\,m$ in the $(\theta_1, \theta_2)$ plane.
We follow binaries having two distinct $\theta_2(x_{\rm i})$ values for
three configurations having  
$\theta_1 (x_{\rm i})$ values that are $ \leq 10^{\circ}$.
We observe that $ \theta_2' > \theta_1' > \iota'$ for all these binary configurations
while small values of these angular variables indicate the influence of efficient tides
during their formation.
}
\label{figure:th1_th2_iota_x0_0}
\end{figure}

In figure ~\ref{figure:th1_th2_iota_x0_0}, we explore
how the  above angular variables  evolve for  $\theta_1(x_{\rm i})$ values lying below $10^{\circ}$.
These configurations can arise in the formation scenarios involving rather efficient tides along with
isotopic or polar supernovae kicks. This should be evident from
the green dots in the $a= 1000\, M$ scatter plots in figures 5 and 6 of \cite{Gerosa13}.
For the plots in figure ~\ref{figure:th1_th2_iota_x0_0},
we let $\theta_1(x_{\rm i})$ to take
 the three values, namely $2^{\circ}, 5^{\circ}$ and $8^{\circ}$.
These $\theta_1$ spin configurations can have
multiple  resonant $\theta_2(x_{\rm i})$ values
and this is due to the presence of two equilibrium curves
at $r=1000\,m$ in the $(\theta_1, \theta_2)$ plane.
The evolution of $\theta_1, \theta_2$ and $\iota$  turned out to be quite
different for resonant $\theta_2(x_{\rm i})$ values lying below and above $90^{\circ}$.
For example, $\iota (x_0)$ and $\theta_1(x_0)$ values are noticeably higher for binaries
having their resonant $\theta_2(x_{\rm i})$ values that are above $90^{\circ}$.
 We note that binaries with $\iota' \leq 10^{\circ}$ can have formation scenarios
involving either isotopic or polar kicks along with efficient tides.
However, binaries with substantially higher $\theta_1'$ and $\theta_2'$
values demand a binary formation channel involving
efficient tides and isotopic supernovae kicks.
This is mainly because such a formation channel naturally allows the less massive spin to lie
in the neighborhood of $180^{\circ}$ from $\vek j_0$ at large orbital separations
(see figure 5 of \cite{Gerosa13}).

\begin{figure}[!ht]
\begin{center}
\includegraphics[width=115mm,height=40mm]{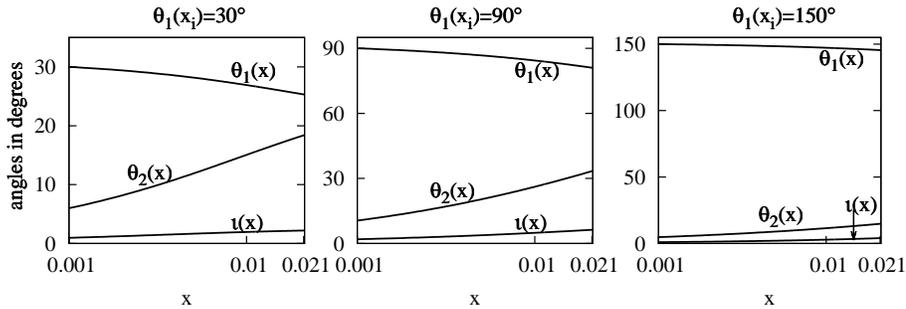}
\end{center}
\caption{Plots for binaries that satisfy Schnittman's  $\Delta \phi=180^{\circ}$
equilibrium configurations at $x_{\rm i}$. The figure follows   
reactive evolution of the angles that specify the orientations of 
$\vek s_1, \vek s_2$ and $\vek k$ from $\vek j_0$ for three spin configurations.
The initial $\theta_1(x_{\rm i})$ values are influenced by \cite{Gerosa13}
and we terminate the evolution at $x_0$.
We observe that $\theta_1' > \theta_2' > \iota'$ while typical 
$\iota (x_0)$ values are of the order of few degrees.
}
\label{figure:Res_th1_th2_i_th12_180}
\end{figure}

\begin{figure}[!ht]
\begin{center}
\includegraphics[width=115mm,height=40mm]{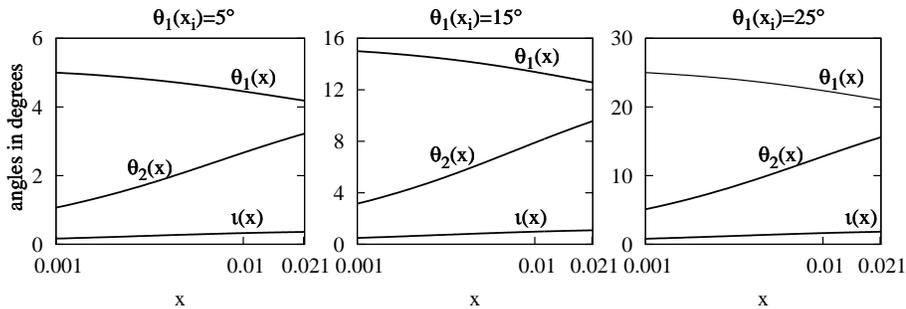}
\end{center}
\caption{ 
Plots, similar to figure~\ref{figure:Res_th1_th2_i_th12_180}, for maximally spinning
BH binaries with $q=11/9$. These spin configurations satisfy the $\Delta \phi=180^{\circ}$
equilibrium solutions at $x_{\rm i}$. 
The three $\theta_1(x_{\rm i})$ values are influenced by \cite{Gerosa13} and are associated 
with the SMR binary formation scenario involving efficient tides and polar kicks.
The orbital angular momentum essentially get aligned with $\vek j_0$ at $x_0$ 
for these binaries.
}
\label{figure:th1_th2_iota_x0_180}
\end{figure}

  We also explored the PN-accurate evolution of binaries influenced by the $\Delta \phi =180^{\circ}$
equilibrium solutions and the results are displayed in figure~\ref{figure:Res_th1_th2_i_th12_180}.
The chosen values of $\theta_1 (x_{\rm i})$ are  $30^{\circ}$, $90^{\circ}$ and $150^{\circ}$.
These initial $\theta_1 $ choices are clearly
influenced by the SMR binary formation channel that also experience efficient tides and 
isotopic kicks \cite{Gerosa13} (see their $a =1000 M$ red scatter plots in figure 5).
These plots 
% in figure~\ref{figure:Res_th1_th2_i_th12_180}
 clearly show that $\theta_1' > \theta_2' > \iota'$ and typical 
$\iota (x_0)$ values are in the $[ 2^{\circ}- 10^{\circ}]$ range.
We also explored  the inspiral dynamics of binaries having $\theta_1 (x_{\rm i})$ lying below 
$30^{\circ}$ and the results are shown in figure~\ref{figure:th1_th2_iota_x0_180}.
Interestingly, we find $\iota'$ values are essentially negligible for initial $\theta_1 $ values below $15^{\circ}$.
It should be noted that these $\theta_1 (x_{\rm i})$ and  $\theta_2 (x_{\rm i})$ values are consistent with 
formation channels involving efficient tides while having both isotropic and polar supernovae kicks. 

  A close inspection of the    
 $\theta_1(x), \theta_2(x)$ and $\iota(x)$ plots of figures~\ref{figure:Res_th1_th2_i_th12}, 
\ref{figure:th1_th2_iota_x0_0}, \ref{figure:Res_th1_th2_i_th12_180} and \ref{figure:th1_th2_iota_x0_180} 
reveals the following interesting point. We infer that  
GWs from inspiraling resonant binaries 
should allow us, in principle, to distinguish between the models of their formation
involving the SMR or RMR scenarios, detailed in \cite{Gerosa13}.
This requires accurate measurements of 
$\theta_1', \theta_2'$ and $\iota'$ 
values that provide the orientations of $\vek s_1, \vek s_2$ and $\vek k$ from $\vek j_0$
when GWs enter interferometric frequency windows.
We find that the SMR scenario binaries 
%that are influenced by $\Delta \phi = \pm \pi$ resonances
tend to have $ \theta_1' > \theta_2' > \iota' $.
The typical values of $\iota'$ are expected to lie below
$10^{\circ}$. 
However, negligible $\iota'$ values are likely for binaries
whose formation scenario involved very efficient tides.
In contrast, in the RMR scenario binaries 
likely to have 
$ \theta_2' > \theta_1' >  \iota' $ and
typical $\iota'$ values are $> 10^{\circ}$.
Binaries with 
non-negligible $\iota'$ values in the range of few degrees
demand RMR formation scenario supplemented by efficient tides.
The above deductions obviously require the crucial inference of \cite{Gerosa13} that 
the SMR and RMR formation channels 
lead to $\Delta \phi = \pm 180^{\circ}$ and $\Delta \phi = 0$ resonances, respectively.
%Additionally, we observe negligible  evolution for 
We also gather from our numerical integrations that  
$\Delta \phi, \gamma$ and $ \dot {\tilde \gamma} \equiv (G\, m/c^3 \, x^{5/2}) \dot {\gamma} $
essentially remain constant
 in the $[10^{-3},0.1]$ $x$ interval for the above two families of equilibrium solutions.
The negligible evolution of these quantities validate Schnittman's observation
that the equilibrium configurations remain
in their associated resonant plane during the inspiral.
Note that the expression for $ \dot {\tilde \gamma}$ allows us to 
follow the variations in the angular part of  
$\dot \gamma$, given by equation~(\ref{eq:gammadot}).

\begin{figure}[!ht]
\begin{center}
$\begin{array}{cc}
\includegraphics[width=65mm,height=60mm]{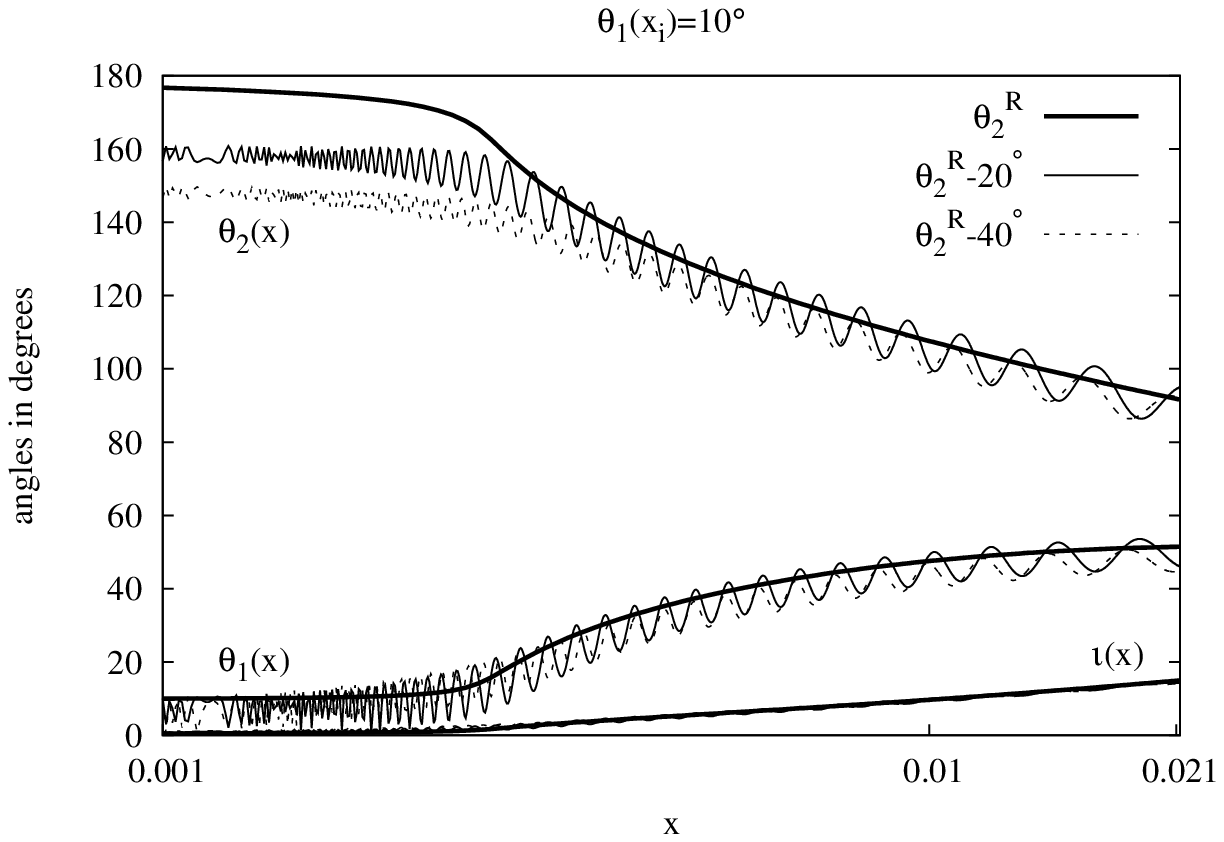}&
\includegraphics[width=65mm,height=60mm]{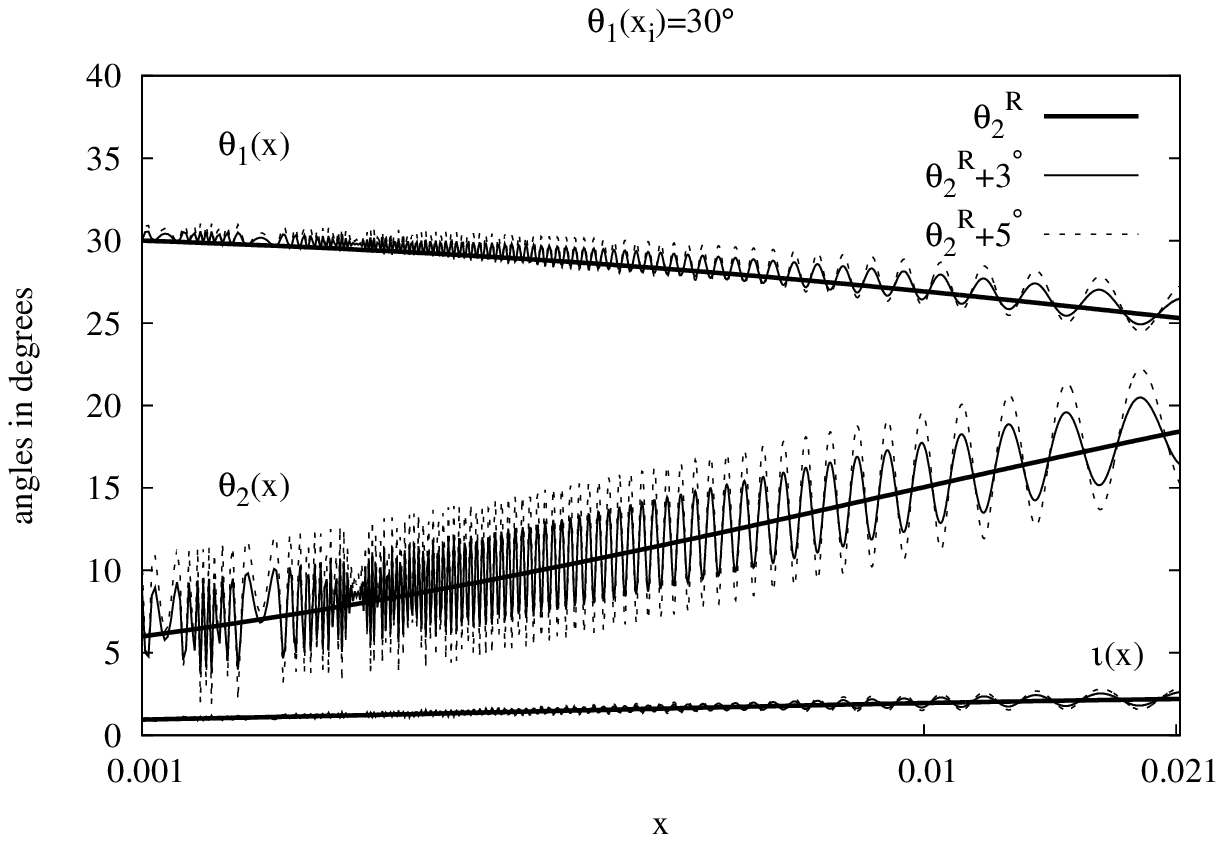}
\end{array}$
\end{center}
\caption{
We depict the evolution of $\theta_1, \theta_2$ and $\iota$ 
for spin configurations
that do not satisfy Schnittman's equilibrium solutions at $x_{\rm i}$.
These non-resonant configurations are constructed by allowing $\theta_2(x_{\rm i})$ 
to be different from their actual resonant values.
The left/right  panel plots are for binaries influenced by the $\Delta \phi =0^{\circ}/ 180^{\circ}$
spin-orbit resonances.
The constraints that distinguish binaries initially in 
the $\Delta \phi =0^{\circ}\,\, {\rm or}\,\, \pm 180^{\circ}$ equilibrium solutions
are also respected by these non-resonant spin  configurations.
During their GW induced inspiral, these configurations get `captured' by the 
spin-orbit resonances and their angular variables librate about the equilibrium 
solutions.
}
\label{figure:th1_th2_iota_nonres_0_pi_x0}
\end{figure}

  It turns out that the above listed constraints also apply for binary configurations that do not
satisfy Schnittman's equilibrium solutions at $x_{\rm i}$.
In figure~\ref{figure:th1_th2_iota_nonres_0_pi_x0}, we present $\theta_1(x), \theta_2(x) $ and $\iota(x)$ plots 
for binaries that do not lie in the 
 $\Delta \phi =0^{\circ}$ and $\Delta \phi = 180^{\circ}$ resonant planes.
This is implemented by choosing $\theta_2(x_{\rm i})$ values that are noticeably different from their 
resonant values at $x_{\rm i}$.
The left panel plots clearly show that $\theta_2'>\theta_1'>\iota'$ and 
we clearly notice the influence of the $\Delta \phi=0^{\circ}$ spin-orbit resonances.
The values of plotted quantities at $x_0$ 
in the right panel plots
are  consistent with 
the SMR formation scenario for these binaries.
We again observe that  $\theta_1'>\theta_2'>\iota'$ as required for binaries influenced by the 
$\Delta \phi=180^{\circ}$ spin-orbit resonances.
We have evolved a number of similar binary configurations 
and the resulting values of $\theta_1, \theta_2 $ and $\iota$ at $x_0$
indeed follow the constraints satisfied by binaries
that lie in the 
 $\Delta \phi =0^{\circ}$ and $\Delta \phi = 180^{\circ}$ resonant planes at $x_{\rm i}$.
Therefore, we conclude that 
GWs from inspiraling spinning  binaries that are influenced by the spin-orbit resonances
should also allow us to distinguish between the models of their formation
involving the SMR or RMR scenarios, detailed in \cite{Gerosa13}.
In appendix, we follow the evolution of spin configurations that do not satisfy the
equilibrium solution at $x_{\rm i}$. This is to probe the behavior of various dynamical variables like 
$\Delta \phi$, $\gamma$, $\dot{\gamma}$ and $\cos^{-1}(\vek s_1 \cdot \vek s_2)$ as such binaries
evolve to $x_0$.

  We also explored the ability of GW measurements to distinguish 
freely precessing binaries from those influenced by the above two families of spin-orbit resonances.
The freely precessing binaries are expected to have $\theta_1(x_{\rm i}) \sim \theta_2(x_{\rm i}) $
as tidal interactions play no significant role during their formation \cite{Gerosa13}.
Moreover, the $\theta_1(x_{\rm i})$ and $\theta_2(x_{\rm i}) $ values can vary essentially between $0^{\circ}$ 
and $180^{\circ}$.
It turns out that these binaries are not affected by the spin-orbit resonances during their GW induced
inspiral from $x_{\rm i}$ to $x_0$.
With the help of our numerical integrations,
we find that  these binaries can mimic
the constraints on the $\theta_1', \theta_2'$ and $\iota'$ values
that are satisfied by the two types of resonant binaries.
However, the $\theta_{12} (x_0)$ values
for freely precessing binaries
 will not satisfy
certain specific relations involving $\theta_1(x_0)$ and $\theta_2(x_0)$ values.
This is important as binaries under the influence of  
$\Delta \phi=0^{\circ}$ ($\Delta \phi=\pm 180^{\circ}$) spin-orbit resonances
are expected to have $\theta_{12} (x_0) \sim \theta_2' - \theta_1' $ 
($ \theta_{12} (x_0) \sim \theta_1' + \theta_2'$) and these 
relations essentially arise from equation~(\ref{Eq_beta}).
Therefore, this additional constraint may be invoked to separate
the freely precessing binaries from the two families of resonant binaries.
To illustrate these points, we provide following two examples.
In the first case, we evolve our typical BH binary having $\theta_1(x_{\rm i}) = \theta_2(x_{\rm i}) =40^{\circ}$
to $x_0$ and the resulting values of $\theta_1, \theta_2, \iota$ and $\theta_{12}$ at $x_0$ are 
around $30^{\circ}, 46^{\circ}, 5^{\circ}$ and $66^{\circ}$, respectively.
The values of the first three angular variables indicate that
the binary can be disguise as one under the influence of
the $\Delta \phi=0^{\circ}$ spin-orbit resonance.
However, the binary's $ \theta_{12} (x_0)$ value does not obey
the required relation, namely $\theta_{12} (x_0) \sim \theta_2' - \theta_1' $ and this
allows us to identify it as a freely precessing one.
In the second example, we let $\theta_1(x_{\rm i}) = \theta_2(x_{\rm i}) =80^{\circ}$
and the resulting $\theta_1', \theta_2', \iota'$ and $\theta_{12}(x_0)$ values are around $91^{\circ}, 36^{\circ}, 14^{\circ}$ and $60^{\circ}$, respectively.
Therefore, it is possible to mistake the configuration as one
influenced by the $\Delta \phi= \pm 180^{\circ}$ spin-orbit resonances, if we restrict
our attention to the values of $\theta_1', \theta_2' $ and $\iota'$.
Fortunately, the fact that  $ \theta_{12} (x_0) \neq \theta_1' + \theta_2'$ allows
us to correctly identify the binary configuration as a freely precessing one.
% {\it We have verified that the above findings are also true for other cases of $\theta_1(x_{\rm i})=\theta_2(x_{\rm i})$.}
These arguments indicate that the accurate measurements of $ \theta_1', \theta_2', \iota'$ and
$\theta_{12} (x_0)$ values are crucial to distinguish the three possible families
of inspiraling comparable mass spinning binaries.

There exists a number of investigations that probed the accuracies
with which GW observations can
estimate various parameters of precessing compact binaries, relevant
for both space
and ground based GW observatories \cite{Lang2006, Klein2009, Lang2011,CHO2013}.
These Fisher information matrix based studies indicate that precessing spins
along with amplitude corrected GW polarization states
 tend to improve measurement accuracies of various intrinsic and
extrinsic parameters that characterize
the inspiral waveform.
The intrinsic parameters include $m,q,\chi_1, \chi_2$ and angular
variables like $\tilde \theta_1$ while
extrinsic parameters refer to the initial phase and angles like $\Theta$.
Very recently, \cite{RO2014} explored measurement accuracies for
precessing BH-NS binaries
with the help of  Markov-chain Monte Carlo simulations.
This detailed study observed that the mass parameters can be better
constrained compared to their
non-precessing counterparts. Additionally, various binary orientation
parameters can be
 measured with reasonable accuracies of the order of few percents
 for fairly loud GW inspiral signals having signal-to-noise-ratio $ \rho \simeq20$.
It will be desirable to pursue a similar study while invoking
our approach to compute $h_{\times, +}(t)$ associated with spinning
compact binaries influenced by spin-orbit resonances.
Such an analysis would allow one to estimate the accuracies with which
aLIGO observations can
estimate the
orientations of $\vek k$, $\vek s_1$ and $\vek s_2$ from $\vek j_0$.
This in turn should allow us to distinguish
between the two possible families of spin-orbit resonances
 in a quantitative manner and to constrain  possible binary formation scenarios.

  In the next section, we provide a prescription to compute $h_{\times,+}(t)$ associated with 
spinning binaries experiencing the spin-orbit resonances that
inspiral through the frequency windows of GW detectors. 
We discuss implications of our approach that includes probing 
the consequence of such binaries not staying in a resonant plane 
when they enter the aLIGO frequency window. We also discuss certain preliminary
data analysis implications of these templates.

\section{GW phasing for compact binaries experiencing the spin-orbit resonances}
\label{Sec_III}

   We adapt an approach, detailed in \cite{GG1}, to accurately model
temporally evolving GW polarization states for inspiraling compact binaries experiencing
the spin-orbit resonances.
In this approach, one invokes $\vek L$ rather than its Newtonian 
counterpart to describe the binary orbits and the two spins are initially specified in a $\vek j_0$-based inertial frame.
This should allow us to incorporate easily various inputs from the previous section 
into the present task.
In what follows, we briefly describe the phasing approach of \cite{GG1} and
we begin by listing the following expressions for $h_{\times}$ and $h_+$, computed using    
the Newtonian accurate expression for the quadrupole moment of the binary.
These expressions, derived in ~\cite{GG1}, read 
\begin{subequations}
\label{eq:hplus_hcross}
\begin{align}
\label{eq_hcross}
h_{\times}|_{\rm Q} &= \frac{2\, G\, \mu\, v^2}{c^4\, R'}\,\Biggl\{(1
-\cos \iota)\, S_{\theta}\,  \sin \iota\, \sin (\alpha- 2 \Phi)   \nonumber \\
&\quad - (1+\cos \iota )\,S_{\theta}\, \sin \iota \, \sin (\alpha + 2\Phi)  \nonumber \\
&\quad -\frac{1}{2}(1+ 2 \cos \iota+ \cos^2 \iota)\, C_{\theta} \sin (2\alpha +2\Phi)  \nonumber \\
&\quad -\frac{1}{2}(1- 2 \cos \iota+ \cos^2 \iota)\, C_{\theta} \sin (2\alpha -2\Phi)  
\Biggr\} \,, \\
\label{eq_plus}
%%%%%%%%%%%%%%%%%%%%%%%%%%%%%%%%%%%%%%%%%%%%%%%%%%%%%%%%%%%%%%%%%%%%%%%%%%%%%%%%%%%%%%%%
h_+|_{\rm Q} &=  \frac{2\, G\, \mu\, v^2}{c^4\, R'}\,\Biggl\{\left(\frac{3}{2}\,\cos^2 \iota
- \frac{3}{2}\right)\, (1-C_{\theta}^2) \cos 2\Phi  \nonumber \\
&\quad - \left(1+\cos \iota\right)\,S_{\theta}\,C_{\theta}\, \sin \iota\, \cos (2\Phi+\alpha)  \nonumber \\
&\quad - \frac{1}{4}( \cos^2 \iota+ 2\,\cos \iota 
+1 )\,(1+C_{\theta}^2)\, \cos (2\alpha + 2\Phi)  \nonumber \\
&\quad - \frac{1}{4}( \cos^2 \iota- 2\,\cos \iota 
+1 )\,(1+C_{\theta}^2)\, \cos (2\alpha - 2\Phi)  \nonumber \\
&\quad-S_{\theta}\, C_{\theta}\, \sin \iota \, \cos \iota \, \cos (\alpha-2 \Phi)    \nonumber \\
&\quad+S_{\theta}\, C_{\theta}\, \sin \iota \cos (\alpha-2\Phi)\Biggr\} \,,     
\end{align}
\end{subequations}
 where $ v^2/c^2, C_{\theta}$ and $S_{\theta}$ 
stand for $ ( G\, m\, \dot \Phi/c^3)^{2/3}, \cos \Theta$ and $\sin \Theta$, respectively.
% = ( G\, m\, \dot \Phi/c^3)^{2/3}$
%while $C_{\theta}$ and $S_{\theta}$ stand for $\cos \theta$ and $\sin \theta$
It should be obvious that
to obtain $h_{\times, +}|_{\rm Q}(t)$, associated with precessing spinning compact binaries
inspiraling along circular orbits, we need to specify how $\iota$, $\alpha$, $\Phi$ and
$\dot{\Phi}$ are varying in time. 
This is achieved by simultaneously solving (numerically) the differential equations 
for the Cartesian components of the fully 2PN-accurate expressions for $\vek s_1$, $\vek s_2$ and $\vek k$,
 given by equations~(\ref{eq:s1_s2_dot}) and (\ref{eq:kdot}),
in the $\vek j_0$-based inertial frame along with
PN-accurate differential equations for $\Phi$ and $x$. 
The differential equation for $\Phi$ is given by
\begin{align}
\label{Eq_phidot}
 \dot \Phi =  \frac{ x^{3/2} }{ ( G\, m/c^3)} - \cos \iota\, \dot \alpha \,,
\end{align}
where we invoked the definition $\omega=v/r$ and
employed the expression for
$\vek v$ in the co-moving triad to derive the above differential equation for $\Phi$ \cite{LK_95,GG1}. 
The effect of GW emission on the above precessional dynamics is 
incorporated by describing how $x$ evolves in time.
We use the following fully 2PN
accurate expression for $\dot{x}$
that includes 2PN-accurate non-spinning, 1.5PN order spin-orbit and 2PN order quadrupole-monopole and
spin-spin self interactions \cite{LK_95,BDIWW,EP98,mvg05,rbk09}.
The resulting expression reads 
 %extracted from \cite{BDIWW,LK_95,EP98,mvg05,rbk09}, given by
\begin{eqnarray}
%\begin{align}
\label{Eq_dxdt2}
\frac{ d x}{dt} &=& \frac{64}{5}\frac{c^3}{Gm}\eta\, {x}^5 
\biggl \{
1+x \left [ -\frac{743}{336}-\frac{11\eta}{4}\right ] +4\pi x^{3/2} \nonumber \\
&&+ \frac{x^{3/2}}{12}\,\Bigl [ (-188\,X_1+75\sqrt{1-4\eta})\,X_1\, \chi_1\, (\vek s_1\cdot \vek k) \nonumber \\
&&+(-188\,X_2-75\sqrt{1-4\eta})\,X_2\,\chi_2\, (\vek s_2 \cdot \vek k)\Bigr ]
\nonumber \\
&&+x^2\,\Bigl[\Bigl ( \frac{34103}{18144}+\frac{13661}{2016}\eta +\frac{59}{18}\eta^2 \Bigr )  \nonumber \\
&&- \frac{1}{48} \eta \chi_1 \chi_2 \,\Bigl ( 247 (\vek s_1 \cdot \vek s_2) -721 (\vek s_1 \cdot \vek k) (\vek s_2 \cdot \vek k)  \Bigr )  \nonumber \\
&& +X_1^2\, \chi_1^2\, \Bigl( \frac{5}{2}\, (3\, (\vek k \cdot \vek s_1)^2-1)+\frac{1}{96}\, (7-(\vek k \cdot \vek s_1)^2)\Bigr)\nonumber \\
 && +X_2^2\, \chi_2^2\, \Bigl( \frac{5}{2}\, (3\, (\vek k \cdot \vek s_2)^2-1)+\frac{1}{96}\, (7-(\vek k \cdot \vek s_2)^2)\Bigr)
  \Bigr]
\biggr \} \,.
%\end{align}
\end{eqnarray}

  Therefore,   we numerically solve eleven coupled
differential equations to obtain $h_{\times,+}|_{\rm Q}(t)$.
These equations include the nine equations for the Cartesian components of
$\dot {\vek k}, \dot {\vek s}_1$ and $\dot {\vek s}_2$, given by equations~(\ref{eq:s1_s2_dot}) and (\ref{eq:kdot})
in the $\vek j_0$-based inertial frame, displayed in figure~\ref{figure:frame}
along with PN-accurate differential equations, namely equations~(\ref{Eq_phidot}) and (\ref{Eq_dxdt2}) for $\Phi $ and $x$.
At every epoch, we extract the values of $\alpha$, $\iota$, 
$\Phi$ and $\dot{\Phi}$ and
hence evaluate the expressions for $h_{\times,+}|_{\rm Q}$, given by equations~(\ref{eq:hplus_hcross}).
This is how we obtain temporally evolving GW polarization states for 
{\it regular} inspiraling precessing compact binaries.
Let us note that we extract 
the angular variables $\iota$ and $\alpha$ from the three Cartesian components of $\vek k$
at every epoch with the help of
$\iota = \cos^{-1} (k_{\rm z})$ and
$ \alpha = \tan^{-1}(k_{\rm y}/k_{\rm x}) $.

   It should be obvious that we  require to impose certain restrictions on the 
angles specifying the initial orientations of $\vek s_1, \vek s_2$ and $\vek k$
at $x_0$  
to obtain $h_{\times,+}|_{\rm Q}(t)$ associated with inspiraling spinning compact binaries
that belong to Schnittman's equilibrium configurations.
These restrictions are imposed in few steps and 
we begin by specifying  $(m_1$, $m_2$, $\chi_1$, $\chi_2)$ values  making 
sure that $m_1\sim m_2$.
% as the resonant binaries are expected to contain comparable masses. We
We also freely specify the values of $(\theta_1$, $\phi_1)$ at $x_0$ and 
these two angles provide the initial orientation of the more massive BH spin $\vek s_1$
at the initial aLIGO frequency
in our $\vek j_0$-based inertial frame.
The requirement that $\gamma$, given by equation~(\ref{eq:gamma}), should be zero allows us to 
let $\phi_2=\phi_1$ at $x_0$ (in other words, we are focusing on the 
$\Delta \phi=0^{\circ}$ equilibrium configurations).
These are the only inputs required to numerically obtain the $\theta_2(x_0)$ value
and this is  achieved, as noted earlier, by equating the following expression
for $\dot{\tilde \gamma}$ to zero. The relevant expression for $\dot{\tilde \gamma}$ reads 
\begin{eqnarray}
\label{Eq_dttgamma}
 \dot{\tilde \gamma} &=& \frac{3}{2}\, \frac{\delta m}{m}\, (\beta -z_1\, z_2) 
 + x^{1/2}\,\biggl [\delta_1\, q\, \chi_1\, (z_2-\beta\, z_1)+ \frac{\delta_2}{q}\, \chi_2\, (\beta\, z_2 - z_1)   \nonumber \\
 &&+ \frac{1}{2}\, X_1^2\, \chi_1\, (-z_2 -2\, \beta\, z_1 + 3\, z_1^2\, z_2)  
 + \frac{1}{2}\, X_2^2\, \chi_2\, (z_1 +2\, \beta\, z_2 - 3\, z_1\, z_2^2) \nonumber \\
 &&+ \frac{3}{2}\, \eta\, (\beta -z_1\, z_2)\, (\chi_1\, z_1 -\chi_2 \, z_2) \biggr ] 
 + \frac{3}{2}\, x \, \biggl[ \eta \, \chi_1 \, \chi_2 (z_1^2-z_2^2)   \nonumber \\
 &&+ X_1^2\, \chi_1^2\, z_1 \, (z_1\, \beta -z_2)+ X_2^2\, \chi_2^2\, z_2 \, (z_1 -\beta\, z_2) \biggr] \,,
\end{eqnarray}
where we invoke equations~(\ref{eq_z1_z2_b_xyz}) for 
$z_1, z_2$ and $\beta$ to express $ \dot{\tilde \gamma}$ in terms of the Cartesian 
components of $\vek k$, $\theta_1, \theta_2, \phi_1$ and $\phi_2$. It should be noted that 
the three Cartesian components of $\vek k$ are provided by equations~(\ref{Eq_ialpha_ini}) and therefore 
the above expression for  $\dot{\tilde \gamma}$
 depends on $\theta_1, \theta_2, \phi_1, \phi_2, m, \eta, \chi_1$ and $\chi_2$ values.
We equate such an expression for $\dot{\tilde \gamma}$ to zero and obtain numerically
$\theta_2(x_0)$ value. 
This approach also allows us to perform the following  internal 
consistency check by obtaining (again) estimates for $\iota$ and $\alpha$ at $x_0$
by using the numerically extracted $\theta_2(x_0)$ value and equations~(\ref{Eq_ialpha_ini}).
We generally use the following 2PN-accurate expression for $L$ at $x_0$
%while computing the Cartesian components of $\vek k$
 in the place of $L_{\rm i}$ in equations~(\ref{Eq_ialpha_ini})
while constructing templates.
The 2PN accurate expression for $|\vek L|$
in terms of $x$, available in ~\cite{BBF}, reads
\begin{equation}
\label{Eq_L2}
L_{\rm 2PN}(x_0)= \frac{G\,m^2\,\eta}{c}\,x^{-1/2}\,
\biggl \{ 1+ x\bigg[\frac{3}{2}+\frac{\eta}{6} \bigg] 
 +x^2\,\bigg[\frac{27}{8}-\frac{19\eta}{8}+\frac{\eta^2}{24}\bigg] \biggr \}  \,.
%\end{align}
\end{equation}

  We are now in a position  to compute 
$h_{\times,+}|_{\rm Q}(t)$ associated with inspiraling compact binaries  
that lie in a resonant plane in the aLIGO frequency window.
We proceed by computing the Cartesian components of $\vek s_1 $ and $\vek s_2$ 
at $x_0$ using the freely specified $(\theta_1, \phi_1)$ values and the above described
$(\theta_2, \phi_2)$
estimates.
The associated Cartesian components of $\vek k$ at $x_0$ arise from equations~(\ref{Eq_ialpha_ini}) while 
invoking the above 2PN-accurate expression for $|\vek L|$ in the place of $L_{\rm i}$.
These initial conditions are invoked 
while numerically solving the eleven 
differential equations, namely equations~(\ref{eq:s1_s2_dot}), (\ref{eq:kdot}), (\ref{Eq_phidot}) and (\ref{Eq_dxdt2}),
% 2PN dx/dt
to obtain the  $h_{\times,+}|_{\rm Q}(t)$ time-series (we let the initial value of $\Phi$ to be zero).
We terminate these numerical integrations when $x$ reaches the value $1/6$
that corresponds to the last stable orbit in the Schwarzschild space-time.

 In what follows, certain preliminary data analysis implications of our above inspiral templates 
for binaries in equilibrium configurations are probed.
This is motivated by the possibility of invoking inspiral templates for  binaries in the  equilibrium  configurations
 to capture GWs from binaries influenced by the spin-orbit resonances.
It should be noted that inspiraling binaries in Schnittman's equilibrium spin configurations
are characterized by essentially 
two angular parameters while one requires four angular parameters 
to fix the initial orientation of the two spins for binaries 
under the influence of 
 the spin-orbit resonances.
This interesting prospect, initially suggested by Schnittman, is probed
by computing the  match ${\cal M}( s,  h)$ involving the expected 
signal waveforms $s(t)$ and the employed template waveforms $h(t)$ \cite{DIS98}. In our $\cal{M}$ computations, $s(t)$ represents 
the inspiral GW signal from binaries influenced by the spin-orbit resonance. The template
waveforms $h(t)$, as expected, model inspiral GWs from binaries in equilibrium configurations as detailed above
(we usually employ the expression for $h_+|_Q$, given by equation (\ref{eq_plus}), to obtain temporally evolving $s(t)$ and $h(t)$). 
To obtain ${\cal M}(s, h)$, we first define the overlap between $s(t)$ and $h(t)$
as 
\begin{equation}
 \mathcal{ O}( s, h ) =   < \hat s, \hat h>  =
\frac{\langle s|h \rangle}{\sqrt{\langle s|s \rangle \, \langle h|h\rangle}} \,,
\end{equation}
where $\hat s$ and $\hat h$ stand for certain normalized GW signal $s(t)$ and the associated template $h(t)$, respectively.
The angular bracket between $ s$ and $ h$ defines certain  noise 
weighted inner product, namely
\begin{equation}
\langle  s |  h \rangle= 4\, {\rm Re}\,  \int_{f_{\rm low}}^{f_{\rm cut}} \, 
\frac{\tilde s^*(f)\, \tilde h^*(f)}{S_{\rm h}(f)} df \,.
\end{equation}
In the above equation $\tilde s(f)$ and $\tilde h(f)$ stand for the Fourier transforms of $s(t)$ and
$h(t)$, respectively, while $S_{\rm h}(f)$ provides 
the one-sided power spectral density (we invoked the 
zero-detuned, high power sensitivity curve of aLIGO \cite{LIGO_2010}). 
In our computations, the upper cut-off
frequency $f_{\rm cut}$ is given by $c^3/(G\, m\, \pi\, 6^{3/2})$ while we let the lower cut-off
frequency $f_{\rm low}$ to be $10$ Hz.
We obtain the match ${\cal M}( s, h)$ by maximizing the $\mathcal{ O}(s, h)$ over 
the time of arrival $t_0$ and the associated phase $\phi_0$:
\begin{equation}
\label{Eq_match}
{\cal M }= \max_{t_0, \phi_0}\, \mathcal{O}(s, h)\,.
\end{equation}
The results of these match computations are displayed in table~\ref{Tab_I}. 
The expected inspiral GW waveforms are from the binaries in certain `near- and far-resonant' configurations for 
four $\theta_1(x_0)$ values: $10^{\circ}, 20^{\circ}, 35^{\circ}$ and $45^{\circ}$.
The near- and far-resonant configurations have $\theta_2(x_0)$ values that differ from their equilibrium 
values by $\pm 5^{\circ}$ and $\pm 20^{\circ}$, respectively.
The match estimates indicate that the near-resonant configurations tend to have $\cal{M}$ values $>0.9$ 
while these estimates are $<0.9$ for far-resonant configurations. 
Additionally, we list the differences in the accumulated phase ($\Delta \Phi$) in aLIGO frequency window 
between the resonant, near- and far-resonant configurations. The $\Delta \Phi$ values, as expected, are 
large for far-resonant configurations and their match numbers are comparatively lower.      
The high $FF$ expectation originates from an investigation that probed the 
`effectualness' of non-precessing spin templates to capture inspiral GWs from comparable mass 
precessing binaries \cite{PA11}. 
The reported very high $FFs$ ($>0.97$) for GWs from a significant fraction of comparable mass 
precessing binaries were attributed to the fact that the precessional effects are less 
influential for such binaries.
The above $0.97$ $FF$ value corresponds to a loss in the event rate not more than $10\%$ of the 
possible sources within the reach of GW detectors.
However, it is rather non-trivial and computationally expensive to pursue similar $FF$ computations involving 
inspiral templates for binaries in equilibrium configurations. This is because the procedure 
involves, in principal, maximization over 
several binary parameters like 
$m, \eta, \chi_1, \chi_2$ and $\theta_1(x_0)$. Finally, 
we note that the listed $\cal{M}$ numbers of table~\ref{Tab_I} are rather insensitive to the employed 
differential equation for $x$. The changes in the $\cal {M}$ estimates were found to be 
less than one part in hundred 
while employing $dx/dt$ that incorporated all the 3.5PN accurate non-spinning contributions.

 \begin{table}[!ht]
%\centering
\caption{Our estimates for the match and the  differences in accumulated phase in the aLIGO frequency window
for the BH binaries
 ($q=11/9, m=30M_{\odot}$) having different values of $\theta_1$ at $x_0$.
The second and third columns are 
for binaries in the near-resonant and far-resonant configurations, respectively.
We let the near- and far-resonant configurations to have $\theta_2(x_0)$ value that are $\theta_2^R\pm 5^{\circ}$ and 
$\theta_2^R\pm20^{\circ}$, respectively. 
The $\Delta \Phi$ (in radians) and $\cal {M}$ estimates associated with the $\theta_2^R+5^{\circ}$ and $\theta_2^R+20^{\circ}$
configurations are shown in the parentheses.
The resonant $\theta_2$ values at $x_0$, denoted by 
$\theta_2^R$, are $21.43^{\circ}$, $41.29^{\circ}$, $66.07^{\circ}$ and $79.11^{\circ}$ for the four 
$\theta_1(x_0)$ values and  $\Delta \phi=0^{\circ}$ equilibrium configurations.
We let $\phi_1=45^{\circ}$ and $\Phi(x_0)=0$.
}
\begin{center}
\begin{tabular*}{0.88\textwidth}{@{\extracolsep{\fill} }c c c c c}
\hline \hline
   $\theta_1(x_0)$                  &  \multicolumn{2}{c}{Near-resonance} &  \multicolumn{2}{c}{Far-resonance} \\  
 \cline{2-4}   \hline
                                        &   $\Delta \Phi$    & $\cal {M}$  & $\Delta \Phi$  & $\cal {M}$ \\ 
                                        \hline
    $10^{\circ}$			&  0.77 (0.95)            & 0.988 (0.918)       & 1.89 (4.80)          & 0.881 (0.860)      \\
    \hline 
    $20^{\circ}$			&  1.52 (1.66)           & 0.963 (0.900)       & 5.09 (7.32)         & 0.852 (0.789)      \\
    \hline
    $35^{\circ}$			&  2.14 (2.18)           & 0.910 (0.921)       & 8.10 (8.76)         & 0.771 (0.760)      \\
    \hline
    $45^{\circ}$			&  2.24 (2.22)            & 0.926 (0.925)       & 8.88 (8.53)         & 0.731 (0.754)     \\
    \hline
    \hline
\end{tabular*}
\end{center}
\label{Tab_I}
\end{table}

  Finally, let us note that it should  be possible to construct inspiral templates for binaries residing
in the resonant planes while invoking a $\vek L_{\rm N}$-based non-inertial triad 
to specify the initial spins.  
This will obviously require us to follow what is summarized in section~\ref{Sec_IIa} 
to obtain these specific spin configurations in an orbital
triad. However, a number of steps are required to evaluate the expressions for 
$ h_{\times}|_{\rm Q}(t)$ and $ h_{+}|_{\rm Q}(t)$ that require
the $\vek j_0$-based inertial frame.
In the first step, the three Cartesian components of the total angular momentum at 
 $x_0$ should be computed.
These components define  two angles, $\theta_j$ and $ \phi_j$,
that specify the orientation of $\vek j_0$ in the $\vek L_{\rm N}$-based 
non-inertial frame.
The second step requires us to rotate the $\vek j_0$, $\vek L_{\rm N}, \vek s_1$ and $\vek s_2$ vectors,
by the following two angles, namely $-\theta_j$ and $ -\phi_j$.
This results in a new Cartesian coordinate system 
where $\vek j_0$ points along the $z$-axis and $\vek L_{\rm N}$ is specified by $(\sin \theta_j, 0, \cos \theta_j)$.
This is the frame where one obtains the temporally evolving $ h_{\times}|_{\rm Q}(t)$ and $ h_{+}|_{\rm Q}(t)$
by simultaneously solving the Cartesian components of $\dot {\vek L}_{\rm N}, \dot {\vek s_1}$ and $\dot {\vek s_2}$ 
along with the PN-accurate differential equations for $\Phi$ and $x$.
It should be noted that this inertial frame is different from our
$\vek j_0$-based inertial frame,  depicted in figure~\ref{figure:frame}.
This is because the $x$ and $y$ axes of these two $\vek j_0$-based inertial frames do not usually coincide.
This should be  evident from the fact that $\vek L_{\rm N}$ at $x_0$ 
is specified only by one angle, namely $\theta_j$
in the new inertial frame. 
Therefore, the resulting resonant plane coincides with the plane where both $\vek N$ and $\vek j_0$ reside.
This implies that $\vek N$, $\vek j_0$, $\vek L_{\rm N}$, $\vek s_1$ and $\vek s_2$ share a common
plane for equilibrium configurations specified in a $\vek L_{\rm N}$-based orbital triad.
Fortunately, the GW phase evolution is not affected by such differences between the two inertial frames.

\section{Conclusions}
\label{Sec_dis_con}

We explored the dynamics of isolated comparable mass spinning compact binaries influenced by 
Schnittman's post-Newtonian spin-orbit resonances 
in an inertial frame associated with $\vek j_0$.
In contrast, it is customary to describe these special equilibrium configurations in a non-inertial
orbital triad \cite{JS,KSB09}.
We argued that accurate GW based estimates of 
the orientations of $\vek s_1, \vek s_2$ and $\vek k$ from 
$\vek j_0$ at $x_0$  should allow us to 
distinguish between the two possible families of spin-orbit resonances.
This should be astrophysically  interesting as  
inspiraling binaries, influenced by 
$\Delta \phi = 0^{\circ}$ 
($\Delta \phi = \pm 180^{\circ}$) spin-orbit resonances, are expected to originate
 from the reverse mass ratio (the standard mass ratio)
formation scenarios \cite{Gerosa13}. 
Therefore, the accurate measurements of $\theta_1', \theta_2'$ and $\iota'$ should, in principle,
allow us to obtain direct observational evidence of  possible
 binary formation channels.
The above deductions also
apply for binaries that do not remain in a resonant plane 
when they become detectable by GW interferometers.
The resonant plane, characterized by either  $\Delta \phi = 0^{\circ}$ or
$\Delta \phi = \pm 180^{\circ}$ restrictions, 
naturally appears in   Schnittman's 
 one parameter family of equilibrium solutions.
It turned out that the two black hole spins and the orbital angular momentum
usually do not lie in such resonant planes during the later stages of binary inspiral.
We emphasized that the accurate GW aided measurements of $ \theta_1', \theta_2', \iota'$ and 
$\theta_{12} (x_0)$ values will be
crucial to distinguish the three possible families 
of inspiraling comparable mass spinning binaries, namely freely precessing binaries
and those influenced by the two types of spin-orbit resonances.
 We also  
developed a prescription to compute the time-domain inspiral templates
for binaries affected  by the spin-orbit resonances.
 We pursued preliminary data analysis implications of such templates by computing the match estimates. 
The aLIGO relevant $\cal {M}$ computations invoked inspiral templates for binaries residing in and
librating around the equilibrium configurations. The resulting match estimates 
point to the possibility that a resonant inspiral template bank may 
 provide the acceptable $FFs\sim 0.97$ 
for inspiral GWs from binaries influenced by spin-orbit resonances.

  It should be interesting to incorporate the higher order spin-orbit and spin-spin contributions 
while constructing inspiral templates for binaries in resonant configurations. At present, this 
is not a straightforward exercise due to the non-availability of next-to-leading order spin-spin 
contributions to $dx/dt$. Note that the next-to-leading order spin-orbit contributions 
to $\dot {\vek s_1}$, $\dot {\vek s_2}$, $\dot {\vek k}$ and $\dot {x}$ are indeed available in~\cite{BBF, FBB}
that are compatible with our equations (\ref{eq:s1_s2_dot}), (\ref{eq:kdot}) and (\ref{Eq_dxdt2}).
In contrast, the next-to-leading order spin-spin contributions to $\dot {\vek s_1}$, $\dot {\vek s_2}$ and $\dot {\vek k}$
require rather detailed manipulations as the associated orbital dynamics, available in \cite{HSH2013}, follow different gauge and  spin supplementary condition. 
However, we do not expect that such higher PN order corrections to 
$\dot {\vek s_1}, \dot {\vek s_2}, \dot {\vek k}$ and $\dot x$ will influence 
our estimates for 
the orientations of $\vek s_1, \vek s_2$ and $\vek k$ from 
$\vek j_0$ at $x_0$.
This is because the present description is sufficient to accurately describe 
the inspiral dynamics of these precessing binaries from $x_{\rm i}$ to $x_0$. 
It will also be interesting to compute the accuracies with which we
can estimate $\theta_1'$, $\theta_2'$ and $\iota'$ values by adapting the detailed analysis presented in \cite{RO2014}.

\ack{We are grateful to 
 Davide Gerosa, Michael Kesden, Richard O'Shaughnessy and Gerhard Sch\"afer for 
helpful discussions and  detailed 
comments.
}

\appendix

\section{Implications of binaries not in spin-orbit resonant configurations} 
\label{appendix}

\begin{figure}[!ht]
\begin{center}
\includegraphics[width=110mm,height=70mm]{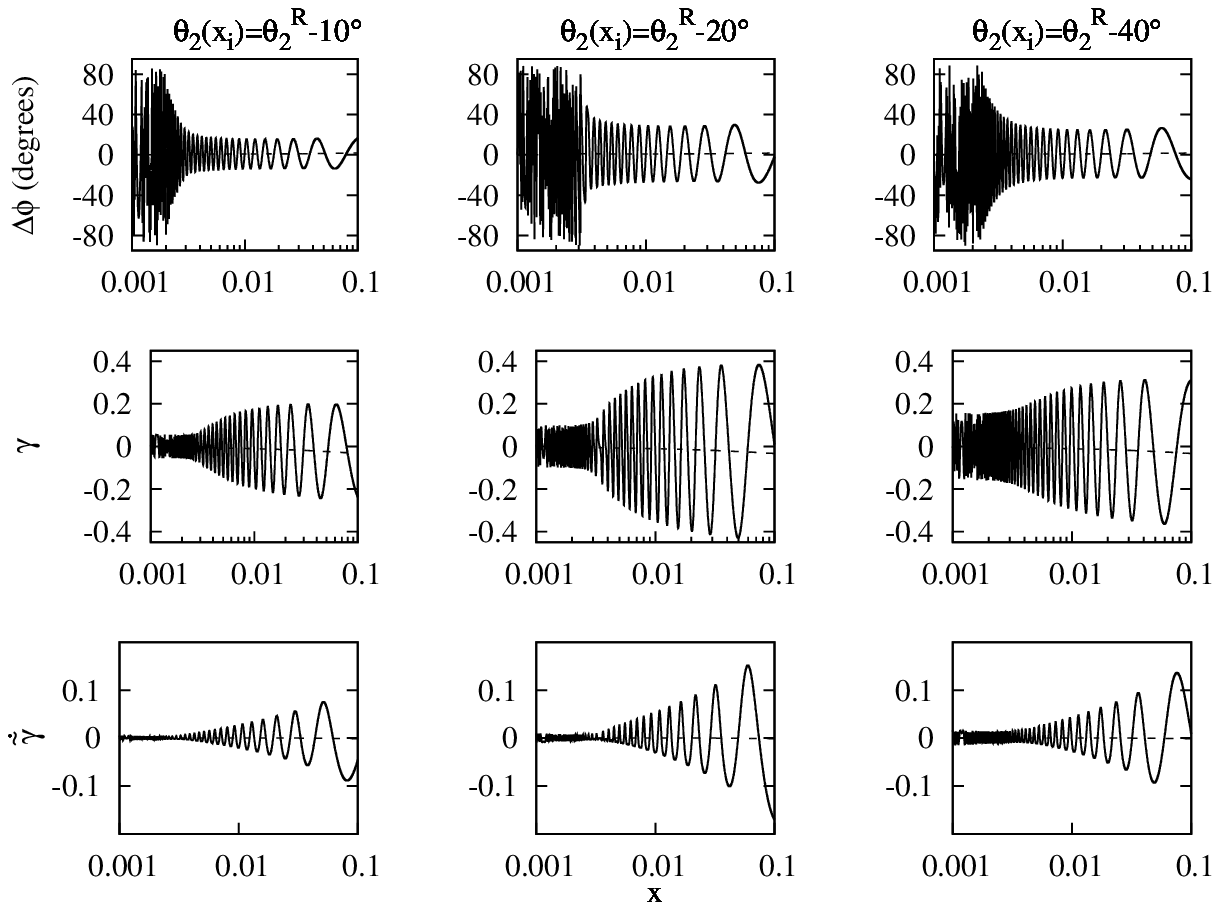}
\end{center}
\caption{ Plots of  $\Delta \phi(x)$, $\gamma(x)$ and $\dot {\tilde \gamma}(x) $
for our typical BH binaries initially not in Schnittman's $\Delta \phi =0^{\circ}$ equilibrium
configurations ($q=11/9$ and the initial dominant spin orientation $\theta_1(x_{\rm i})=20^{\circ}$).
The three columns are for three different initial $\theta_2$ values, namely $\theta_2^R(x_{\rm i})-10^{\circ}$, 
$\theta_2^R(x_{\rm i})-20^{\circ}$, $\theta_2^R(x_{\rm i})-40^{\circ}$
where the resonant $\theta_2$ value, namely $\theta_2^R(x_{\rm i})=173.71^{\circ}$.
These quantities follow secular evolution  for 
binaries that reside in the $\Delta \phi=0^{\circ}$ resonant plane as shown by the dashed line plots. 
 The non-negligible $\gamma$ and $\dot{\tilde \gamma}$ evolution indicates 
that $\vek k, \vek s_1$ and $\vek s_2$ do not share a plane during the late stages of binary inspiral.
These quantities remain essentially unchanged for binaries that reside in the $\Delta \phi=0^{\circ}$
 resonant plane (dashed line).
The amplitudes of $\gamma$ and $\dot{\tilde \gamma}$ evolution
turned out to be noticeably lower for binaries having $\theta_1(x_{\rm i}) \leq 10^{\circ}$.
This is because 
such spin configurations are likely to lie 
in the close neighborhood of another ($\theta_1, \theta_2$) equilibrium solution
at $x_{\rm i}$.
}
\label{figure:NonRes_delphi_g_gdot_20}
\end{figure}

 In what follows, we probe the behavior of various dynamical variables for spin configurations
that do not force $\gamma$ and $\dot {\gamma}$ to be zero at $x_{\rm i}$.
It was noted that such configurations usually approach and librate around the  equilibrium
configurations  with steadily decreasing amplitudes during the inspiral as evident from figure~5 in \cite{JS}.
This prompted Schnittman to suggest that during the inspiral 
generic spin configurations can approach
the equilibrium configurations and eventually get locked into the spin-orbit resonances.
The steadily decreasing amplitude of $\Delta \phi$ oscillations implies that
the orbital and spin angular momenta will eventually lie in a plane and 
this was termed as `resonant plane locking' in \cite{JS,Gerosa13}.
In what follows, we explore the ability of gravitational radiation reaction to 
force the orbital and spin angular momentum vectors to lie in a plane 
for binaries influenced by the spin-orbit resonances.
This is done by following the 
evolution of $\Delta \phi, \gamma$ and $ \dot {\tilde \gamma}$ in the $[10^{-3}-0.1]$ $x$ interval 
while choosing $\theta_1(x_{\rm i})$ value to be $20^{\circ}$.
The initial  $\theta_2$ values at $x_{\rm i}$ differ by 
$10^{\circ},20^{\circ} $ and $40^{\circ}$ from the actual 
resonant value (the resonant $\theta_2$ value at $x_{\rm i}$  
being $173.71^{\circ}$).
In figure~\ref{figure:NonRes_delphi_g_gdot_20}, we
display the results of our PN-accurate evolution of these configurations.
The plots for $\Delta \phi$  show that initially
these binaries precess freely through 
a large range in $\Delta \phi$ and the gravitational radiation reaction forces a substantial 
reduction in these wild oscillations. 
However, $\Delta \phi$ librates about $\Delta \phi=0^{\circ}$ with essentially constant amplitude 
during the substantial part of the inspiral to $x_{\rm f}$
and not with a steadily decreasing amplitude as noted in ~\cite{JS}.
Moreover, the plots for $\gamma$ and $\dot{\tilde \gamma}$ are oscillatory 
and $\gamma$ librates around the resonant value, namely $\gamma =0$, 
with roughly constant and non-negligible amplitudes
as these binaries inspiral to $x_0$.
%
%during the most part of binary inspiral.
%These oscillations have roughly constant and non-negligible amplitudes
%after these binaries inspiral to $x_0$.
We have verified that 
the temporal oscillations in $\gamma$ and $ \dot {\tilde \gamma}$
are such that they do not simultaneously approach zero at any epoch during their inspiral
from $x= 10^{-2}$ to $x_{\rm f}$. 
Therefore, it is reasonable to infer that the unit vectors along 
the black hole spins and orbital angular momentum in these 
binaries do not lie in a resonant plane,
 characterized
by negligible values of $\gamma$ and $ \dot {\tilde \gamma}$,
when their GWs become detectable by aLIGO.

      Let us emphasize that evolution of comparable mass spinning binaries, not in the resonant configurations at 
$x=10^{-3}$, indeed gets influenced by the spin-orbit resonances
and therefore the two spins do not precess freely towards the end of the inspiral.
However, such resonances are not very efficient in forcing the two spins 
and the orbital angular momentum to share a common plane during the late inspiral 
as evident from the non-negligible values of $\gamma$ in our figures.
We gather from a number of similar numerical experiments that 
the amplitude of $\Delta \phi(x)$ oscillations can become small 
for spin configurations where  $\vek s_1$ orientations from $\vek j_0$ at $x_{\rm i}$
are  $ < 10^{\circ}$. 
These $\theta_1 (x_{\rm i})$ configurations can have 
multiple $\theta_2(x_{\rm i})$ values 
that satisfy Schnittman's $\Delta \phi=0^{\circ}$ equilibrium solution.
This is due to the presence of two equilibrium curves
at $r=1000\,m$ in the $(\theta_1, \theta_2)$ plane (see the left panel plots of figure~\ref{figure:th1_th2}).
This forces many more $\theta_2 (x_{\rm i})$ values, lying in the range $[0^{\circ}- 180^{\circ}]$,
to approach the neighborhood of $\Delta \phi=0^{\circ}$ equilibrium solutions.
These configurations turned out to have 
$\gamma (x) $ values that are $< 0.1$ when they inspiral to $x_0$.

\begin{figure}[!ht]
\begin{center}
\includegraphics[width=95mm,height=70mm]{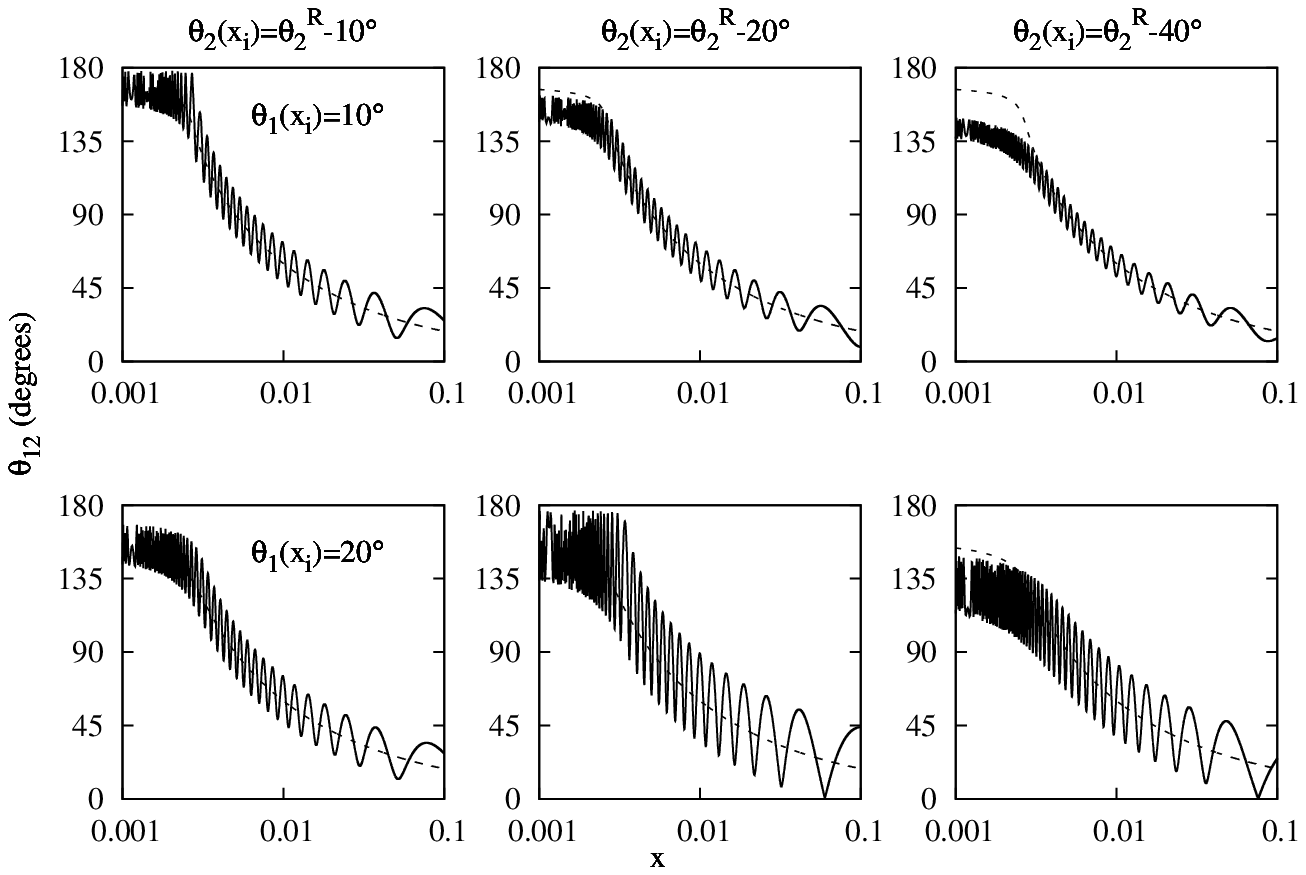}
\end{center}
\caption{ Semi-log plots for the $\theta_{12}=\cos^{-1}(\vek s_1\cdot \vek s_2)$ evolution for binary configurations
having $q=11/9$. 
We consider  three non-resonant binary 
configurations having their $\theta_2(x_{\rm i})$ values differing from their $\theta_2^R(x_{\rm i})$ values by $10^{\circ}$, $20^{\circ}$
and $40^{\circ}$. The two rows are
for two different $\theta_1(x_{\rm i})$ values, namely $\theta_1(x_{\rm i})=10^{\circ}$ and $20^{\circ}$.
The oscillatory nature of the $\theta_{12}(x)$ plots is
consistent with the figure~6 in ~\cite{JS}.
The spin-alignment also happens for non-resonant configurations and their $\theta_{12}$ evolution clearly 
follow how $\theta_{12}$ vary for the associated equilibrium solution, depicted by the dotted lines.
}
\label{figure:NonRes_th12}
\end{figure}

   Finally, we would like to point out that our numerical integrations are fully consistent with 
those presented in \cite{JS,KSB09,BKS12}. This is demonstrated by probing 
the ability of spin-orbit resonances to 
align the two spins with each other for spin configurations lying in the neighborhood of 
the $\Delta \phi=0^{\circ}$  equilibrium solutions.
% during the late inspiral.
The spin alignment is clearly visible in the  plots of figure~\ref{figure:NonRes_th12}
where we follow $\theta_{12}(x)$ evolutions for two initial $\theta_1$ values that are $\leq 20^{\circ}$.
The plots along the two rows are for the two specific $\theta_1(x_{\rm i})$ values, namely 
$10^{\circ}$ and $20^{\circ}$ while we vary $\theta_2(x_{\rm i})$ values along 
the three columns.
The chosen $\theta_2(x_{\rm i})$ values differ from their actual resonant values 
by $10^{\circ},20^{\circ} $ and $40^{\circ}$ respectively (the
resonant $\theta_2(x_{\rm i})$ values for these two initial 
$\theta_1$ values are $176.75^{\circ}$ and $173.72^{\circ}$, 
respectively).
We clearly observe substantial reductions in the $\theta_{12}(x)$ values as these binaries spiral in to $x_{\rm f}$
and it  essentially leads to the spin-alignment.
However,  $\theta_{12}(x)$ evolution is oscillatory as we move away from the equilibrium 
configurations. The amplitude of $\theta_{12}(x)$ librations about their equilibrium values 
depend on both the initial 
$\theta_1$ values and the departure of $\theta_2(x_{\rm i})$ from its resonant value.
The oscillatory $\theta_{12}(x)$ evolution is also 
consistent with the sinusoidal projections in figure~6 of \cite{JS}.
Therefore, it is reasonable to state that our numerical integrations are consistent 
with the results of \cite{JS,KSB09}.

\end{document}